\documentclass[nofootinbib,aps,showpacs,preprintnumbers,amsmath,amssymb,eqsecnum,floats]{revtex4}

\usepackage{graphics}
\usepackage{graphicx}
\usepackage{color}
\usepackage{dcolumn}

\newcommand{\dst}{\displaystyle}
\newcommand{\pa}{\partial}

\def\np{({\bf n}\cdot{\bf p})}
\def\pp{{\bf p}^2}
\def\ppp{({\bf p}^2)}

\def\hn{\widehat{H}_{\text{N}}}
\def\hi{\widehat{H}_{\text{1PN}}}
\def\hii{\widehat{H}_{\text{2PN}}}
\def\hiii{\widehat{H}_{\text{3PN}}}
\def\hiv{\widehat{H}_{\text{4PN}}}

\begin{document}

\title{Gravitational waves from black hole binary inspiral and merger:\\
The span of third post-Newtonian effective-one-body templates}

\author{Thibault Damour}
\affiliation{Institut des Hautes Etudes Scientifiques,
91440 Bures-sur-Yvette, France}
\author{Bala R.\ Iyer} 
\affiliation{Raman Research Institute, Bangalore 560 080, India}
\author{Piotr Jaranowski}
\affiliation{Institute of Theoretical Physics, University of Bia{\l}ystok,
Lipowa 41, 15-424 Bia{\l}ystok, Poland}
\author{B.\ S.\ Sathyaprakash}
\affiliation{Department of Physics and Astronomy, Cardiff University,
Cardiff, CF2 3YB, U.K.}

\date{\today}

\begin{abstract}

We extend the description of gravitational waves emitted by binary black holes
during the final stages of inspiral and merger by introducing in the third
post-Newtonian (3PN) effective-one-body (EOB) templates seven new
``flexibility'' parameters that affect the two-body dynamics and gravitational
radiation emission.  The plausible ranges of these flexibility parameters,
notably the parameter characterising the fourth post-Newtonian effects in the
dynamics, are estimated.  Using these estimates, we show that the currently
available standard 3PN bank of EOB templates does ``span'' the space of signals
opened up by all the flexibility parameters, in that their maximized mutual
overlaps are larger than 96.5\%.  This confirms the effectualness of 3PN EOB
templates for the detection of binary black holes in gravitational-wave data
from interferometric detectors.  The possibility to drastically reduce the
number of EOB templates using a few ``universal'' phasing functions is
suggested.

\end{abstract}


\maketitle

\section{Introduction}
\label{IntroSec}

Current theoretical understanding, stemming both from general relativity and
astrophysics, places black hole binaries at the top of the list of candidate
sources for the interferometric gravitational-wave detectors that are nearing
the completion of their construction phase.  On the one hand, black hole
binaries are by far sources whose dynamics (early inspiral and late time
quasi-normal mode ringing to a large extent, late inspiral, plunge and merger to
a lesser extent) is better understood than other sources, such as supernovae or
relativistic instabilities in neutron stars, so that it is possible to construct
reasonably good template waveforms to extract signals out of noise.  On the
other hand, astrophysical rate estimates of black hole binary coalescences,
though not known accurately, have a range whose upper limit is large enough to
expect a few mergers per year within a distance of 150 Mpc \cite{glpps}.

The merger phase of binaries consisting of two $15M_\odot$ black holes takes
place right in the heart of the LIGO-VIRGO-GEO sensitivity band giving us the
best possible picture of this highly non-linear evolution.  Eventually, when
detectors reach good sensitivity levels, one hopes to learn experimentally about
this strong gravity regime which has been a subject of intense analytical and
numerical studies for more than a decade.  
In the meantime, what is needed is a set of model waveforms or templates that
describe the dynamics close to the merger phase accurately enough so that only a
small fraction $(<10\%)$ of all events will go undetected.

Two general viewpoints are possible to reach this goal.  A maximalist one or a
minimalist one.  The {\it maximalist} viewpoint consists in enlarging as much as
is conceivable (using in a ``democratic'' way all available methods in the
literature for treating binary coalescence) the bank of filters, with the hope
that even the methods which appear a priori less reliable than others might, by
accident, happen to describe a good approximation to the ``real'' signal.  This
is the viewpoint taken by Buonanno, Chen and Vallisneri in their careful,
and detailed analysis in Ref. \cite{BCV02}, on the basis of which they
advocate expanding the net by using a multiparameter template family able to
approximate most of the results of the conceivable analytical methods.  
However, this ``democratic''  attitude comes at a cost
that calls for an alternate strategy that we explore here.  The
problem with this method is that it leads to a dramatic increase in the total
number of templates from $\sim 25$ templates to $\sim 10^4$ templates; which has
the bad consequence that it leads a larger false alarm rate.  (These estimates
of the number of templates are those obtained, as in \cite{BCV02}, by
``dividing'' the parameter space by the local span of the template at the
minimal match.  This is an underestimate because it neglects boundary effects.
For instance, a more realistic estimate of the number of the third 
post-Newtonian effective-one-body templates would be $\sim 150$ rather than 25.)

By contrast, we advocate here a {\it minimalist} viewpoint consisting in (i)
focussing exclusively on the best available analytical description, and (ii)
generalizing this description by adding several parameters that describe ``new
directions'' corresponding to physical effects not perfectly modelled by this
description.  The most important of these directions are the effects due to
higher post-Newtonian (PN) effects, not yet calculated, but known to exist.  A
thorough study of the robustness of our preferred description against the
inclusion of currently unknown effects should allow an informed and judicious
covering of the parameter space of interest without overtly expanding the size
of the total bank of templates.

The best available analytical description at present is, in our opinion, the
``effective-one-body'' (EOB) approach proposed and constructed for non-spinning
bodies at second post-Newtonian (2PN) order by Buonanno and Damour
\cite{BD99,BD00}, extended to third post-Newtonian (3PN) order by Damour,
Jaranowski and Sch\"afer \cite{DJS00} and generalized to spinning bodies by
Damour \cite{TD02}.  On the one hand, the EOB maps, using the Hamilton-Jacobi
formalism, the real ``conservative'' (in the absence of radiation reaction)
dynamics of two bodies with masses $m_1$ and $m_2$ into an EOB problem of a test
particle of mass $\mu\equiv{M\eta}$ (where $M\equiv{m_1+m_2}$ and
$\eta\equiv{m_1m_2/M^2}$), moving (essentially) in an effective background
metric $g_{\mu\nu}^{\rm eff}$ which is a deformation of the Schwarzschild metric
with deformation parameter $\eta$.  Further, by supplementing the above dynamics
by an additional radiation-reaction force obtained from a Pad\'e resummation of
the gravitational-wave flux, it allows for the first time the possibility to go
beyond the adiabatic approximation and to analytically discuss the transition
from inspiral to plunge and the subsequent match to merger and ringing. 
The implications of the EOB templates for data analysis of binary black holes
were explored in \cite{DIS3} where it was shown that the signal-to-noise ratio
(SNR) is significantly enhanced relative to the usual PN templates due to
inclusion of the plunge signal.  The EOB formalism does also provide initial
dynamical data (position and momenta) for two black holes at the beginning of
the plunge to be used in numerical relativity to construct gravitational data
like metric and its time derivative and evolve Einstein's full equations through
the merger phase.

The analytical prediction of the EOB method (including spin) for invariant
functions were compared to numerical results based on the helical Killing vector
approach \cite{BGG02} for circular orbits of corotating black holes by Damour,
Gourgouhlon and Grandcl\'ement \cite{DGG02} and shown to agree remarkably well.
The agreement was robust against choices of resummation of the EOB potential and
improved with the PN order.  Recently, Buonanno, Chen and Vallisneri
\cite{BCV02} made a detailed and exhaustive comparison of all currently
available waveforms for non-spinning binary black holes resulting from different
approximations.  This study showed (among other results) that EOB models are
more reliable and robust than other non-adiabatic models.
 
Recently Blanchet \cite{B02a} has made a comparison of the straightforward PN
predictions with \cite{BGG02} and shown that at 3PN order they are as close to
the numerical results as the resummed approaches.  While it is indeed
interesting to note this closeness of the results derived from one particular
non-resummed 3PN function (the ``energy function'' $E(\omega)$ for circular
orbits) to
the numerical results (and to the  EOB ones), we still do not see any way yet by 
which, as the Hamiltonian $H({\bf r},{\bf p})$ does in the EOB approach, the 
bare PN results for the `` circular energy function'' $E(\omega)$ can be used to 
define templates beyond the adiabatic approximation.
As Ref.\ \cite{BD00} has shown the importance of going beyond the adiabatic
approximation in describing the smooth transition between the inspiral and
the plunge, and as \cite{DIS3,BCV02} has shown how significant was the contribution
of the plunge to the SNR, we consider that the EOB waveforms
are the best, currently available, analytical templates for binary
black hole coalescences.

In this paper, we wish to strain to extremes the flexibility of the EOB
formalism by tugging it in directions where it can be theoretically pulled and
by locating directions it is most likely to yield under deformations by
unmodelled effects (including higher order PN effects).  We shall introduce
seven ``flexibility'' directions i.e.\  seven new flexibility parameters.
Then we will investigate the ``span'' of the original (3PN) EOB templates in the
space of waveforms opened up by our extension of the EOB waveforms into seven
new ``flexibility'' directions.  By {\it span} of a given bank of
templates, we mean the region of signal space which is well modelled by some
template in the bank, i.e.\  the set of signals $S$ such that the maximized
overlap of $S$ with {\it some} template $T$ is larger than 0.965 \footnote{When
we don't have a perfect template to capture a signal then the effective distance
up to which a detector could have ideally seen goes down. Suppose with the use of
the correct template an antenna could detect sources at a distance $D.$
If the best overlap we can achieve with the true signal is $f$ then that
distance drops to $fD$ and the new rate of events would be 
proportional to $(fD)^3.$  In other words, the fractional decrease in the
number of events is $(1 - f^3).$ By demanding that $(1-f^3) < 0.1$ (i.e. a
loss of no more than 10\% of all potential events) we get $f \simeq 0.965.$}.
Among our
seven flexibility parameters, some like $b_5$ represent higher order (fourth
post-Newtonian, 4PN) corrections in the dynamics, some like $\theta$, the
arbitrariness in the best available 3PN gravitational-wave flux calculation due
to incompleteness of the Hadamard partie finie regularization, and others like
$c_P$ denote a parameter used to factor the gravitational-wave flux in
\cite{DIS1} and accelerate the convergence of the Pad\'e approximants to the
numerical flux in the test-mass case.  Further, the current development of EOB
has made, at several stages, specific {\it choices} of representation of various
physical effects and, as in any analytical construction, the choices were the
{\it simplest} that one could apparently make.  We then explored the effects
induced by a {\it modification} of these simple choices, i.e.\  the
consideration of new versions of EOB, characterised by different parameter
values reflecting other allowed more complex choices.  These comprise the
remaining four parameters
and include $z_2$($\zeta_2$) a parameter appearing in the effective Hamiltonian
$H_{\rm eff}$ of the EOB, $f_{\rm NonAdiab}$, a parameter to modify the simplest
treatment of non-adiabatic effects in the current version of EOB templates;
$f_{\rm NonCirc}$, a parameter to modify the simplest treatment of non-circular
effects in the current version of EOB templates and finally $f_{\rm transition}$
to allow the possibility that the transition between plunge and ringing may
occur at frequencies different from that assumed in the simplest EOB model.  The
seven parameters $(b_5,\,\theta,\,c_P,\,\zeta_2,\,f_{\rm NonAdiab},\,f_{\rm
NonCirc},\,f_{\rm transition})$, are referred to as the {\it flexibility}
parameters.  Varying them and testing the change of the physical predictions
under reasonable variation of these parameters is a way of probing the overall
robustness of the EOB framework.  What we specifically investigate
is whether a bank of standard 3PN EOB templates is sufficient to represent all
plausibly relevant extended family of waveforms generated by these new
flexibility parameters.

In the context of this investigation our waveforms will be parameterized by two
sets of parameters:  (a) The first set consists of the usual intrinsic and
extrinsic parameters entering the construction of standard waveforms, such as
the masses of the component objects and their spins, some reference phase, etc.,
denoted collectively by $p_k,$ $k=1,\ldots, K.$ In this work we shall only deal
with non-spinning point-particles in the {\it restricted} PN approximation
\cite{cutleretal93a} which requires $K=4$ with $p_{1,2}=m_{1,2}$ denoting the
masses of the two bodies, $p_3=t_{\rm ref}$ a reference time (related to the
instant of coalescence), and $p_4=\Phi_0$ the phase of the wave at the reference
time.  (b) The second set consists of the {\it flexibility parameters}
introduced above.  We shall denote these flexibility parameters as $\pi_a,$
$a=1,\dots,A,$ with ($A=7$)
\begin{eqnarray}
\pi_1 & = & b_5,\ \  \pi_2=\theta,\ \  \pi_3=c_P,\ \  \pi_4=z_2(\zeta_2), \ \ 
\nonumber\\
\pi_5 & = & f_{\rm NonAdiab},\ \  \pi_6=f_{\rm NonCirc},\ \  
\pi_7=f_{\rm transition}.
\end{eqnarray}

For studies of the span of a bank of templates of the kind we propose to do in 
this paper, it is helpful to introduce the notions of a standard or {\it  
fiducial template} and its associated variant or {\it flexed signal} constructed 
by turning on one of the flexibility parameters.
The term {\it fiducial template} is used to represent a waveform constructed in 
a certain approximation and at a given PN order with a fixed set of values of 
the unknown parameters introduced above. In this paper, our fiducial template 
will be the standard EOB waveform (see Sec.\ \ref{EOBSec}) at the 3PN order with 
the flexibility parameters all set to zero:
\begin{equation}
\text{Fiducial Template} = h(t; p_k, \pi_b=0, \forall b) \equiv T(t; p_k).
\end{equation}
In contrast, the associated {\it flexed signal} will again be the EOB waveform 
at 3PN order with {\it all but one} of the  flexibility parameters $\pi_a$ set 
to zero:
\begin{equation}
\pi_a\text{-Flexed Signal} = h(t; p_k, \pi_a\ne 0, \pi_b=0, \forall \ b\ne a)
\equiv  S(t; p_k, \pi_a).
\end{equation}
In other words, in our test of robustness we {\it do not allow all the seven
parameters to vary simultaneously}.  Such a variation would lead to a formidably
high dimensional parameter space which is computationally impossible to
investigate at the moment.  Rather, our aim is to study the effect of each
flexibility parameter independently and to gauge the extent to which our
standard fiducial template waveform can mimic the changes brought about by the
flexibility parameters $\pi_a$ by a mere variation of the intrinsic parameters
$p_k$.  Such a systematic study allows us to isolate and identify the most
important unknown physical effects, and decide if it is necessary to introduce
the corresponding flex parameter as an additional parameter in search templates
used in the detection of gravitational waves from black-hole binaries.

As in earlier work, our main tool for measuring the span of a bank of templates
is the {\it overlap} of a standard fiducial template with a given flexed signal.
(See Sec.\ref{sec:Overlap} for the definition of the overlap.) In this study,
we use two measures of ``good overlaps'':  {\it faithfulness} and {\it
effectualness} \cite{DIS1}.  A {\it template} is said to be {\it faithful} if
its overlap with a {\it flexed signal} waveform of {\it exactly the same
intrinsic parameter} values is larger than 0.965 (after maximization over the
extrinsic parameters).  It is expected that faithful
templates are also good at estimating source parameters although this is not
guaranteed to be the case.  A {\it template} is said to be {\it effectual} if
its overlap with a {\it flexed signal} waveform, {\it maximized over all the
intrinsic and extrinsic parameters}, 
is larger than 0.965.  Obviously, every faithful template is necessarily
effectual but not all effectual templates are faithful.  Note that the notions
of faithfulness and effectualness might depend on the particular flexibility
direction which is explored:  While a template waveform could be faithful with
respect to the $\pi_1$-flexed signal, it might only be effectual with respect to
the $\pi_2$-flexed one and neither with respect to the $\pi_3$-flexed one.

Implementing the above analysis we conclude that {\it the standard 3PN EOB
templates are effectual with respect to all flexibility parameters} introduced
in this study including the parameter $b_5$ characterising the 4PN dynamical
effects.  In other words, the {\it span} of the bank of 3PN EOB templates is
large enough to cover the space of signals described by the physically plausible
ranges of the seven flexibility parameters considered here.  No additional extra
parameters are required in the detection templates to model the more complex
choices possible in the EOB approach at the 3PN level or the dominant dynamical
effects at the 4PN order.  In particular, there is no need to increase the total
number of templates beyond the level required for the standard 3PN EOB
templates.

\section{Third post-Newtonian dynamics and  energy flux}
\label{3pnDynamicsSec}

The conservative dynamics of binary systems in the PN approach has now been 
determined to 3PN accuracy. Two independent calculations, one based on the 
canonical ADM approach together with the standard Hadamard partie finie 
regularization for the self-field effects \cite{3PNADM1,3PNADM2} 
and the second, a 
direct 3PN iteration of the equations of motion 
in harmonic coordinates supplemented by an extended Hadamard partie finie 
regularization \cite{3PNBF,BFHad,ABF01} agree that the 3PN dynamics and 
consequently conserved quantities like energy are fully determined except for 
one arbitrary parameter called $\omega_s$ in the ADM approach and $\lambda$ in 
the harmonic coordinates related by,
\begin{equation}
\lambda = -\frac{3}{11}\omega_s - \frac{1987}{3080}.
\end{equation}
The Hadamard regularization of the self-field of point particles used in
\cite{3PNADM1,3PNADM2,3PNBF,BFHad,ABF01,BIcm} has the serious drawback of violating
the gauge symmetry of perturbative general relativity (diffeomorphism
invariance), and thereby of breaking the crucial link between Bianchi identities
and equations of motion.  This explains why the Hadamard-based works
\cite{3PNADM1,3PNADM2,3PNBF,BFHad,ABF01,BIcm} were unable to fix the parameter
$\omega_s$.  Recently, Damour, Jaranowski and Sch\"afer \cite{DJS01} have
proposed to use a better regularization scheme, one which respects the gauge
symmetry of perturbative general relativity: {\it dimensional regularization}.  
They
have implemented this improved regularization scheme, which led them to a unique
determination \cite{DJS01} of the parameter $\omega_s$, namely $\omega_s=0$,
corresponding to $\lambda=-1987/3080$.  Thus the conservative dynamics is
completely determined to 3PN order within the ADM approach using dimensional
regularization.  Though it will be interesting to reconfirm the value of
$\lambda$ by other treatments, we believe that the result of \cite{DJS01} is
trustable especially in view of the obtention there, by the same regularization
method, of the unique Poincar\'e-invariant momentum-dependent part of the
Hamiltonian.  Thus, for all applications including data analysis, there is no
arbitrariness in 3PN dynamics, and consistent with this, in this paper we set
$\omega_s=0$ or equivalently $\lambda=-1987/3080$.

On the other hand, the gravitational-wave energy flux from binary systems has
been computed using the multipolar post-Minkowskian approach \cite{MPM} in
harmonic coordinates and Hadamard regularization to 3.5PN accuracy.  Unlike at
earlier orders \cite{2PN} the instantaneous and hereditary contributions do not
remain isolated.  At 3PN order, in addition to the instantaneous terms, the
tails-of-tails and tail-squared terms also contribute.  Fortunately, they have
been computed by Blanchet \cite{btail} who has also computed the tail
contribution at 3.5PN order.  The gravitational wave energy flux contains the
3PN-accurate time derivative of the mass quadrupole moment leading to a specific
$\lambda$ dependence which as explained earlier is now known since $\omega_s$ is
computed.  However, the incompleteness of the Hadamard regularization introduces
additional arbitrary parameters in the mass quadrupole moment leading to three
new undetermined parameters that combined into the unique quantity $\theta$ in
the circular energy flux.  Unfortunately, up to now no alternate regularization
or calculations without regularizations exist that provide the value of
$\theta$.  Thus, one has to reckon with this arbitrary parameter in the
templates that one constructs and the best one can do is estimate its
implications for data analysis of inspiraling compact binaries as in
\cite{BCV02} or in the present work.

The expression for the 3PN energy function \cite{3PNADM1,3PNADM2,3PNBF} and
3.5PN flux function \cite{BFIJ02,BIJ02} and the resulting 3PN and 3.5PN
coefficients in various phasing formulas discussed in \cite{DIS3} are summarised
in \cite{DIS4}.  Though they are the basis of the present analysis, they are not
reproduced here for reasons of brevity and we refer the reader to \cite{DIS4}
for those expressions.

In this work, we have used the 3PN-accurate flux function because the standard
near-diagonal Pad\'e of the 3.5PN flux function involves a spurious pole in the
physically relevant range of variation of $v$.  We leave to future work an
investigation of alternative Pad\'e's of the 3.5PN flux free of such spurious
poles.  In view of the numerical smallness of the 3.5PN contribution to the
flux, we expect no significant change in our physical conclusions.

\section{Transition from inspiral to merger---the 3PN effective-one-body model}
\label{EOBSec}

The starting motivation of the EOB approach is to try to capture in a small
number of numerical coefficients the essential invariant PN contributions from
among the plethora of terms that exist in the complete PN expansion
of the binary's equations of motion, in the
belief that many of these terms are gauge artefacts and hence irrelevant.
It is also strongly motivated by the need to look for an analytic route to go
beyond the adiabatic approximation which breaks down before the last stable
orbit.  We recall that the standard PN treatments
based on invariant functions $(E(\omega),\,{\cal F}(\omega))$
are limited by the adiabatic approximation (and cannot describe the
transition to plunge), while the treatments based on the direct
use of (non-resummed) PN-expanded equations of motion are 
unreliable \cite{BCV02} because of poor convergence of the straightforward
PN-expanded equations of motion.

As shown in \cite{BD00}, at 2PN order the mapping to EOB is eventually unique
(when imposing some general requirement).  The waveform, the equations governing
the evolution of the orbital phase and the initial conditions to integrate them
through the plunge are discussed in \cite{BD00} and were used in \cite{DIS3} to
construct 2PN EOB templates and investigate their performance.  At 3PN order, on
the other hand, the situation is more involved.  When requiring that the
relative motion be equivalent to geodesic motion in some effective metric, there
are more constraints than free parameters in the energy map and effective
metric.  This led \cite{DJS00} to an extension of the 2PN EOB construction
(non-geodesic motion) involving a larger variety of choices.
In Ref.\ \cite{DJS00} the following generalized 3PN EOB Hamiltonian was 
introduced: 
\begin{equation}
\label{effH1}
\widehat{H}_{\text{eff}}({\bf r},{\bf p})
= \sqrt{ A(r) \left[ 1 + {\bf p}^2 + \left( \frac{A(r)}{D(r)} - 1 \right) \np^2 
+ \frac{1}{r^2} \biglb( z_1\,\ppp^2 + z_2\,\pp\np^2 + z_3\,\np^4 \bigrb)  
\right]} \,,
\end{equation}
where the functions $A(r)$ and $D(r)$ are given by the components of the  
effective spherically symmetric metric $g_{\mu\nu}^{\text{eff}}$: 
$A(r)=-g_{00}^{\text{eff}}(r)$ and $D(r)/A(r)=g_{rr}^{\text{eff}}(r)$ (they also 
depend on the parameters $z_1$ and $z_2$; see below).
Here ${\bf r}$ and ${\bf p}$ denote the (scaled) canonical 
coordinates of the effective dynamics, $r\equiv{|{\bf r}|}$,
${\bf n}\equiv{\bf r}/r$; $r$ is dimensionless, being scaled by $G M$.
The effective position vector ${\bf r}$ is linked \cite{BD99,DJS00}
to the relative position
vector ${\bf x}_1  - {\bf x}_2$ of the two holes in ADM coordinates by a  
post-Newtonian expansion which starts as: 
${\bf r} = ({\bf x}_1  - {\bf x}_2)/(G M)  + {\cal O}( c^{-2})$.

The parameters $z_1$, $z_2$, and $z_3$ are arbitrary, but subject to the  
constraint
\begin{equation}
\label{zcon1}
8z_1 + 4z_2 + 3z_3 = 6(4-3\eta)\eta \,,
\end{equation}
which forbids the geodesic choice $z_1=z_2=z_3=0$ of the 2PN EOB, 
but allows the  minimally 
non-geodesic choice, $z_1=z_2=0$ and $z_3=2(4-3\eta)\eta$.
[See, however, below our discussion of non-minimal choices
as flexibility directions].

In spherical coordinates $(r,\theta,\phi)$, restricting the motion to the 
equatorial plane $\theta=\pi/2$, the Hamiltonian Eq.~(\ref{effH1}) can be 
written as
\begin{equation}
\label{effH2}
\widehat{H}_{\text{eff}}(r,p_r,p_\phi;z_1,z_2,z_3) = \sqrt{ A(r;z_1) \left\{
1 + A(r;z_1)D(r;z_1,z_2)^{-1}\, p_r^2 + \frac{p_{\phi}^2}{r^2}
+ Z(r,p_r,p_\phi;z_1,z_2,z_3) \right\} } \,, 
\end{equation}
where
\begin{equation}
\label{Z}
\dst Z(r,p_r,p_\phi;z_1,z_2,z_3) \equiv \frac{1}{r^2} \left[
z_1 \left(p_r^2 + \frac{p_\phi^2}{r^2}\right)^2
+ z_2 \left(p_r^2 + \frac{p_\phi^2}{r^2}\right) p_r^2 + z_3\, p_r^4 \right] \,.
\end{equation}
The functions $A(r;z_1)$ and $D(r;z_1,z_2)^{-1}$ depend on  $z_1$ and $z_2$:
\begin{subequations}
\begin{eqnarray}
\label{A} 
A(r;z_1) &=& 1 - \frac{2}{r} + \frac{2\eta}{r^3} + \frac{a_4(z_1)}{r^4} \,,
\\[2ex] 
\label{D} 
D(r;z_1,z_2)^{-1} &=& 1 + \frac{6\eta}{r^2}
+ \frac{2(26-3\eta)\eta-7z_1-z_2}{r^3} \,,
\end{eqnarray}
\end{subequations}
where
\begin{equation}
a_4(z_1) \ = \ \left( \frac{94}{3} - \frac{41}{32}\pi^2 \right) \eta - z_1 \,.
\label{a4}
\end{equation}

The (scaled) 3PN {\em EOB-improved real} Hamiltonian is the following function 
of the EOB 
Hamiltonian Eq.~(\ref{effH2}): 
\begin{equation}
\label{realHam} 
\widehat{H}_{\text{real}} \ = \frac{1}{\eta} \
\sqrt{1 + 2\eta(\widehat{H}_{\text{eff}}-1)} \,.
\end{equation}
The equations of motion have the form of the usual Hamilton equations: 
\begin{subequations}
\begin{eqnarray}
\label{Heq1} 
\frac{dr}{dt} &=& \frac{\partial \widehat{H}_{\text{real}}}{\partial p_r}, 
\\[2ex] 
\label{Heq2} 
\frac{d\phi}{dt} &=& \frac{\partial \widehat{H}_{\text{real}}}{\partial p_\phi}, 
\\[2ex] 
\label{Heq3} 
\frac{dp_r}{dt} &=& - \ \frac{\partial \widehat{H}_{\text{real}}}{\partial r}, 
\\[2ex] 
\label{Heq4} 
\frac{dp_\phi}{dt} &=& \widehat{\cal F}_\phi \,, 
\label{eq:dphibydt}
\end{eqnarray}
\end{subequations}
where $\widehat{\cal F}_\phi$ is the $\phi$ component of the damping force.

\subsection{Pad\'e approximants of $A$}

The  straightforward PN expansion of the function $A$, in terms of the variable 
$u\equiv{1/r}$, reads:
\begin{equation}
\label{ATay}
A(u) = 1 - 2u + 2\eta u^3 + a_4(\eta) u^4 + a_5(\eta)u^5 + {\cal O}(u^6) \,.
\end{equation}
To improve the convergence of the PN expansion \eqref{ATay} we introduce the 
following sequence of Pad\'e approximants of $A$ \cite{DJS00}:
\begin{subequations}
\label{ef11}
\begin{eqnarray}
A_{\text{1PN}}(u) &\equiv& 1-2u \,,
\\[2ex]
A_{\text{2PN}}(u) &\equiv&
\frac{1-\left(2-\frac{1}{2}\eta\right)u}{1+\frac{1}{2}\eta u+\eta u^2} \,,
\\[2ex]
A_{\text{3PN}}(u) &\equiv&
\frac{2(4-\eta)+\biglb(a_4(\eta)-16+8\eta\bigrb)u}
{2(4-\eta)+\biglb(a_4(\eta)+4\eta\bigrb)u+2\biglb(a_4(\eta)+4\eta\bigrb)u^2
+4\biglb(a_4(\eta)+\eta^2\bigrb)u^3} \,,
\\[2ex]
A_{\text{4PN}}(u;a_5) &\equiv&
\frac{ \biglb(16-8\eta-a_4(\eta)\bigrb)
- \biglb(32-24\eta-4a_4(\eta)-a_5(\eta)\bigrb)u }{d_{\text{4PN}}(u;a_5)} \,,
\end{eqnarray}
\end{subequations}
where
\begin{eqnarray}
d_{\text{4PN}}(u;a_5) &\equiv&
\biglb( 16-8\eta-a_4(\eta) \bigrb)
+ \biglb( 8\eta +2 a_4(\eta)+a_5(\eta) \bigrb) (u+2u^2)
+ 2\biglb( 8\eta^2+(4+\eta)a_4(\eta)+2a_5(\eta) \bigrb) u^3
\nonumber\\[2ex]&&
+ \biglb( 16\eta^2+a_4(\eta)^2+8\eta a_4(\eta)+(8-2\eta)a_5(\eta) \bigrb) u^4 
\,.
\end{eqnarray}
Reference \cite{DGG02} has studied some variants of the specific Pad\'e choices
made in Eqs.\ (\ref{ef11}) and found that they had very little effect on
physical quantities down to the last stable orbit.  Therefore, we shall not
include below, among our flexibility directions, the ones corresponding to
different choices of definition of $A$ function.  We shall always define it
using the $P^1_n$-type Pad\'e used above.  While the 3PN-level coefficient
$a_4(\eta)=b_4\eta$ is known when using $\omega_s=0$ \cite{DJS01} [see Eq.\
(\ref{b4}) below], the 4PN-level coefficient $a_5(\eta)$ introduced here is
unknown.  Its possible values will be discussed below.

\subsection{3PN EOB adiabatic initial data}

The initial dimensionless frequency $\widehat{\omega}_0$ depends on the initial 
frequency $f^0_{\text{GW}}$ of the gravitational wave and the total mass $M$ of 
the 
binary system:
\begin{equation}
\widehat{\omega}_0 = \frac{GM\pi f^0_{\text{GW}}}{c^3} \,.
\label{td12}
\end{equation}
In the following equations $A$ and $D$ are treated as functions of $u$,
${\dst A'\equiv {dA}/{du}}$.
The initial value of ${\dst r_0\equiv {1}/{u_0}}$ one obtains solving 
numerically equation:
\begin{equation}
\widehat{\omega}_0 = {\dst u^{3/2} \sqrt{\dst \frac{\dst -\frac{1}{2}A'}{1
+2\eta\left(\dst \frac{A}{\sqrt{A+\frac{1}{2}uA'}}-1\right)}} } \,.
\label{eq:omegahatVSu}
\end{equation}
The initial momenta are then obtained from equations:
\begin{subequations}
\begin{eqnarray}
\label{pphi}
p_\phi^0 &=& \left. \sqrt{-\frac{A'}{u\left(2A+uA'\right)}}\, 
\right\vert_{u=u_0} \,,
\\[2ex]
\label{prad}
p_r^0 &=& \left.  
\frac{u\left(2A+uA'\right)A'D}{A\left[2u(A')^2+AA'-uAA''\right]}\, 
\right\vert_{u=u_0}
\frac{\widehat{\cal F}_\phi(\widehat{\omega}_0)}{\widehat{\omega}_0} \,.
\end{eqnarray}
\end{subequations}

\subsection{3PN EOB light ring}

The light ring coordinate ${\dst u_{\text{light ring}}={1}/{r_{\text{light 
ring}}} }$ is the solution of equation
\begin{equation}
A(u) + \frac{1}{2}uA'(u) = 0 \,.
\label{eq:rlightring}
\end{equation}
See Ref.\ \cite{BD00} for a discussion of the physics associated with the light
ring within the EOB framework.
[See also \cite{LBLR} for alternative views based on the non-resummed 3PN
energy function $E(\omega)$.]

\section{Estimating the effect of unknown physics}
\label{sec:unknown physics}

Our main goal is to vary the flexibility parameters within a certain range
motivated by physical arguments to be discussed in Secs.\ 
\ref{b5Sec}--\ref{transitionSec} and to determine the degradation
caused by such a variation on the {\it overlap}
of flexed waveforms with fiducial waveforms.  If the degradation is small then
we further explore the maximum extent to which the parameters can be
meaningfully varied so that the effectualness (see Sec.\ \ref{sec:faithfulness}
for a definition) still remains more than 96.5\%.  We limit the range of
variation of each parameter so that the effective potential, binding energy and
energy flux remain regular and meaningful for values of the parameter in that
range.  The natural range over which the parameters are expected to vary is
summarized in the first row of Table~\ref{table:max range}.

\subsection{Higher order PN dynamics parameter $b_5$}
\label{b5Sec}

In terms of the inverse $u\equiv{1/r}$ of the radial effective coordinate $r$ 
the PN expansion (\ref{ATay}) of the function $A(u)=-g_{00}^{\text{eff}}(u)$ can 
be written as:
\begin{equation}
\label{Au1}
A(u) = 1 - 2\, u + b_3\, \eta\, u^3 + b_4\, \eta\, u^4
+b_5 \eta\, \left[1 + {\cal O}(\eta)\right]u^5+ {\cal O}(u^6)\,, 
\end{equation}
where
\begin{subequations}
\begin{eqnarray}
b_3 &=& 2 \,,
\\[2ex]
\label{b4}
b_4 &=& \frac{94}{3}-\frac{41}{32}\pi^2 \,.
\end{eqnarray}
\end{subequations}
We have included in Eq.\ (\ref{Au1}) the information that for the 2PN ($\propto 
u^3$) and 3PN ($\propto u^4$) levels there has been rather miraculous 
cancellations to leave only terms linear in $\eta$. Moreover, we also know that 
to all orders (starting from 2PN) the terms $\propto\eta^0$ vanish.
We expect 
that the terms linear in $\eta$
in the higher PN coefficients dominate over the nonlinear ones. 
In particular, we expect that in the 4PN coefficient $a_5(\eta)=
b_5\eta +{\cal O}(\eta^2)$, the term ${\cal O}(\eta^2)$ can be neglected.
Finally, we shall work under the simple assumption $a_5(\eta)=b_5\eta$,
i.e.\ with 
\begin{equation}
\label{Au2}
A(u) = 1 - 2\, u + b_3\, \eta\, u^3 + b_4\, \eta\, u^4
+ b_5\, \eta u^5 + {\cal O}(u^6) \,.
\end{equation}
The main aim of this subsection will be to estimate the plausible order of
magnitude of the 4PN coefficient $b_5$.

We can start to guess a plausible range of values of $b_5$ by the following
reasoning.  It is plausible to expect that the function $A(u)$ be a meromorphic
function of $u$ (or, at least, be close to a meromorphic function).  The growth
with $n$ of the Taylor coefficients $b_n$ of a meromorphic function is
determined by the location of the nearest singularity in the complex plane of
the function $A(u)$.  If the nearest singularity is located at $u=1/b$, the
Taylor coefficients $b_n$ of $A(u)$ behave, when $n$ increases, roughly
proportionally to $b^n$.  We can always parametrise this behaviour (without loss
of generality) as
\begin{equation}
\label{pow1}
b_n \simeq k \ b^{n-3} \,.
\end{equation}
Using Eq.\ (\ref{pow1}) we can deduce $b$ and $k$ from $b_3$ and $b_4$:
\begin{subequations}
\label{bc}
\begin{eqnarray}
k &\simeq& b_3 \ = \ 2 \,,
\\[2ex]
b &\simeq& b_4/b_3 \ \simeq \ 9.3 \,.
\end{eqnarray}
\end{subequations}
This yields the guess
\begin{equation}
b_5 \ \simeq \ 170 \,.
\end{equation}

Note that the values of $b$ and $k$ from Eqs.\ (\ref{bc}) gives also a 
``prediction'' for $b_2$ which is
\begin{equation}
b_2 \ \simeq \ k/b \ \simeq \ 0.2 \,.
\end{equation}
This type of value is small enough not to make a physical difference from the
exact value $b_2=0$, and moreover it is compatible with the fact that the
detailed calculation of $b_2$ gives a cancellation of the type $b_2=0.25-0.25$
(see text below and Table \ref{table:bnk}), each term being indeed of order 0.2.
This is consistent with a power-law growth of the typical contributions entering
$b_n$.

A first guess is therefore that $b_5$ is positive (because this is true for
$b_3$ and $b_4$) and smaller than 200 (in round numbers).  To go beyond this
guess we studied in detail the various PN contributions to the successive $b_n$,
with the aim of detecting a pattern.  We explain in detail our study in the
remainder of the subsection.

\subsubsection{$E=E(\lowercase{j})$ for circular orbits}

We work here with the Hamiltonian describing the real relative motion of the two 
bodies in their center-of-mass reference frame (the superscript NR denotes a 
``non-relativistic'' Hamiltonian, i.e.\ the Hamiltonian without the rest-mass 
contribution). It reads
\begin{equation}
\label{ham}
\widehat{H}^{\text{NR}}\left({\bf r},{\bf p}\right)
= \widehat{H}_{\rm N}\left({\bf r},{\bf p}\right)
+ \frac{1}{c^2} \widehat{H}_{\rm 1PN}\left({\bf r},{\bf p}\right)
+ \frac{1}{c^4} \widehat{H}_{\rm 2PN}\left({\bf r},{\bf p}\right)
+ \frac{1}{c^6} \widehat{H}_{\rm 3PN}\left({\bf r},{\bf p}\right)
+ \frac{1}{c^8} \widehat{H}_{\rm 4PN}\left({\bf r},{\bf p}\right),
\end{equation}
where
\begin{equation}
\label{rpdef}
{\bf r} \equiv \frac{{\bf x}_1-{\bf x}_2}{GM},\quad
{\bf p} \equiv \frac{{\bf p}_1}{\mu} = -\frac{{\bf p}_2}{\mu},\quad
\widehat{H}^{\rm NR} \equiv \frac{H^{\rm NR}}{\mu}.
\end{equation}
Let us note that in this subsection (and only here) ${\bf r}$ and ${\bf p}$ 
denote the canonical coordinates of the real (relative) two-body dynamics.

Circular motion is defined through the condition
\begin{equation}
\label{np=0}
{\bf n}\cdot{\bf p} = 0,
\end{equation}
where ${\bf n}\equiv{\bf r}/r$ and $r\equiv|{\bf r}|$. Under the condition 
Eq.~(\ref{np=0}) the Hamiltonians $\hn$ through $\hiv$ have the structure
\begin{subequations}
\begin{eqnarray}
\hn &=& \frac{\pp}{2} - \frac{1}{r} \,,
\\[2ex]
\hi &=& \sum_{k=0}^2 h_1^k \frac{\ppp^k}{r^{2-k}} \,,
\\[2ex]
\hii &=& \sum_{k=0}^3 h_2^k \frac{\ppp^k}{r^{3-k}} \,,
\\[2ex]
\hiii &=& \sum_{k=0}^4 h_3^k \frac{\ppp^k}{r^{4-k}} \,,
\\[2ex]
\hiv &=& \sum_{k=0}^5 h_4^k \frac{\ppp^k}{r^{5-k}} \,.
\end{eqnarray}
\end{subequations}

The momentum squared $\pp$ can always be decomposed as
\begin{equation}
\pp \equiv \np^2 + ({\bf n}\times{\bf p})^2 = p_r^2 + \frac{j^2}{r^2},
\end{equation}
where $p_r \equiv {\bf n}\cdot{\bf p}$ and
\begin{equation}
j \equiv |{\bf j}| \,, \quad
{\bf j} \equiv \frac{\bf J}{\mu GM} = {\bf r}\times{\bf p} \,.
\end{equation}
Here ${\bf J}$ is the conserved total angular momentum of the binary system in 
the center-of-mass reference frame.

The Hamiltonian (\ref{ham}), after replacing $\pp$ by $j^2/r^2$, becomes a 
function of $j$ and $r$ only. This function has to fulfill, by virtue of  the
equations of motion (for circular motion $p_r=0$), the condition
\begin{equation}
\label{rj}
\frac{\pa \widehat{H}^{\text{NR}}(j,r)}{\pa r} = 0 \,.
\end{equation}
Equation (\ref{rj}) gives the link between $r$ and $j$ along circular orbits.
We have iteratively solved Eq.\ (\ref{rj}) for $r$ as a function of $j$, and 
have substituted this into the Hamiltonian (\ref{ham}). We thus have obtained 
the relation, valid along circular orbits, between the center-of-mass energy 
$E\equiv\widehat{H}^{\text{NR}}\left({\bf r},{\bf p}\right)$ of the system and 
the system's angular momentum $j$. To simplify displaying this relation we show 
it with the coefficients $h_1^k$ of the 1PN Hamiltonian $\hi$ replaced by their 
explicit general relativistic values. Then the formula reads
\begin{eqnarray}
\label{Ej}
E &=& -\frac{1}{2j^2} \left\{
1 + \frac{1}{4}(9 + \eta) \frac{1}{j^2}
+ \left( 16 - 2 \sum_{k=0}^3 h_2^k \right) \frac{1}{j^4}
+ \left( 152 - 24\,\eta - 8 \sum_{k=0}^3 (k+3)\,h_2^k - 2 \sum_{k=0}^4 h_3^k
\right) \frac{1}{j^6}
\right.\nonumber\\[2ex]&&
\phantom{-\frac{1}{2j^2} \left\{\right.}
+ \left( 1700 - 584\,\eta + 36\,\eta^2
+ 4 \sum_{k=0}^3 (k+3)(3\,\eta-27-4\,k)\,h_2^k
+ \left( \sum_{k=0}^3 (k+3)\,h_2^k \right)^2
\right.\nonumber\\[2ex]&&\left.\left.
\phantom{-\frac{1}{2j^2} \left\{+\left(\right.\right.}
- 8 \sum_{k=0}^4 (k+4)\,h_3^k - 2 \sum_{k=0}^5 h_4^k
\right) \frac{1}{j^8}
\right\} \,.
\end{eqnarray}

\subsubsection{EOB potential $A(\lowercase{u})$ calculated from 
$E=E(\lowercase{j})$}

In the  effective-one-body approach the real ``non-relativistic'' energy $E$ is 
the 
following function of the effective-one-body radial potential $W_j(u)$:
\begin{equation}
\label{ereal}
E = \frac{1}{\eta} \left\{
\sqrt{1 + 2\eta \left(\sqrt{W_j(u)} - 1\right)} - 1 \right\} \,,
\end{equation}
where
\begin{equation}
\label{Wjdef}
W_j(u) = A(u) \left( 1 + j^2 u^2 \right) \,.
\end{equation}
The function $A(u)$ has a perturbative expansion in $u$:
\begin{equation}
\label{Apert}
A(u) = 1 + a_1\,u+ a_2\,u^2 + a_3\,u^3 + a_4\,u^4 + a_5\,u^5 \,.
\end{equation}

Along circular orbits the effective radial potential $W_j(u)$ attains its
minimal value,
\begin{equation}
\label{Wjmin}
\frac{\pa}{\pa u} W_j(u) = 0 \,.
\end{equation}
We have iteratively solved Eq.\ (\ref{Wjmin}) for $u$ as a function of the small
parameter $1/j^2$ and have substituted this relation into the right-hand side of
formula (\ref{ereal}).  Next we have again expanded the right-hand side of
(\ref{ereal}) in $1/j^2$.  In such a way we have obtained the relation $E=E(j)$
predicted by the general EOB function (\ref{Apert}) in which the coefficients at
different powers of $1/j^2$ depend on the numbers $a_n$ entering the function
$A(u)$.  By comparing these coefficients with the respective coefficients of the
expansion (\ref{Ej}) we are able to (iteratively) express $a_n$ in terms of the
coefficients $h_n^k$ of the Hamiltonian (\ref{ham}).

After this matching between the generic Hamiltonian (\ref{ham}) and 
the guessed EOB expression (\ref{Apert}),
each of the numbers $a_n$ can be represented as a sum of terms which depend on 
the coefficients of the different PN Hamiltonians. E.g., $a_2=a_{20}+a_{21}$,
where $a_{20}$ depends only on the coefficients of the Newtonian Hamiltonian 
$\hn$ and  $a_{21}$ depends on the coefficients of the Newtonian $\hn$ {\em and}
the 
1PN $\hi$ Hamiltonians.  More generally, we have
\begin{equation}
a_n = \sum_{k=0}^{n-1} a_{nk} \,,
\end{equation}
where $a_{n0}$ depends only on $\hn$, $a_{n1}$ depends on $\hn$
and $\hi$, $a_{n2}$ depends on $\hn$, $\hi$ and $\hii$, etc.
The values of the different coefficients $a_{nk}$ are as follows (here again,
to 
simplify formulas, the coefficients $h_1^k$ of the 1PN Hamiltonian $\hi$  
have  been 
replaced by their general relativistic values)
\begin{subequations}
\begin{eqnarray}
a_{10} &=& -2 \,,
\\[2ex]
a_{20} &=& \frac{1}{4} \left(9 + \eta\right) \,,
\\[2ex]
a_{21} &=& -\frac{1}{4} \left(9 + \eta\right) \,,
\\[2ex]
a_{30} &=& \frac{1}{16} \left(-27 - 12\,\eta + \eta^2\right) \,,
\\[2ex]
a_{31} &=& \frac{1}{16} \left(-67 + 30\,\eta + \eta^2\right) \,,
\\[2ex]
\label{a50}
a_{32} &=& 2 \sum_{k=0}^3 h_2^k \,,
\\[2ex]
a_{40} &=& \frac{1}{64} \left(54 + 72\,\eta - 11\,\eta^2 + 2\,\eta^3\right) \,,
\\[2ex]
a_{41} &=& \frac{1}{64} \left(-1973 + 1301\,\eta - 55\,\eta^2 - \eta^3\right) 
\,,
\\[2ex]
a_{42} &=& \sum_{k=0}^3 (3 - \eta + 8\,k)\,h_2^k \,,
\\[2ex]
a_{43} &=& 2 \sum_{k=0}^4 h_3^k \,,
\\[2ex]
a_{50} &=& \frac{1}{1024}
\left( -243 - 1080\,\eta + 210\,\eta^2 - 104\,\eta^3 + 21\,\eta^4 \right) \,,
\\[2ex]
a_{51} &=& \frac{1}{1024}
\left( -201017 + 250392\,\eta - 46630\,\eta^2 + 536\,\eta^3 + 15\,\eta^4
\right) 
\,,
\\[2ex]
a_{52} &=& \frac{1}{8} \sum_{k=0}^3
\left( 207 + 384\,k + 128\,k^2 - (213 + 128\,k)\,\eta + 7\,\eta^2 \right) h_2^k
- \sum_{k=1}^3 k\,h_2^k \sum_{k=0}^3 (6+k)\,h_2^k \,,
\\[2ex]
a_{53} &=&  \sum_{k=0}^4 (5 - \eta + 8\,k)\,h_3^k  \,,
\\[2ex]
\label{a54}
a_{54} &=& 2 \sum_{k=0}^5 h_4^k \,.
\end{eqnarray}
\end{subequations}

Because the Hamiltonians from Newtonian through 3PN are completely known, the 
coefficients $a_1$ through $a_4$ are  fully known. They read
\begin{subequations}
\begin{eqnarray}
a_1 &=& -2 \,,
\\[2ex]
a_2 &=& 0 \,,
\\[2ex]
a_3 &=& 2\,\eta \,,
\\[2ex]
a_4 &=& \left(\frac{94}{3}-\frac{41}{32}\pi^2\right)\eta \,.
\end{eqnarray}
\end{subequations}
Let us note that $a_3$ and $a_4$ are both proportional to $\eta$.
Many remarkable cancellations occurred to cancel the terms proportional
to $\eta^2$ and $\eta^3$.
 As for the 4PN-level coefficient $a_5=\sum a_{5k}$
its first three partial contributions $a_{51}$, $a_{52}$, $a_{53}$
are known, but its last coefficient $a_{54}$ is unknown.
$a_5$ is a polynomial of order at most 4 in $\eta$ with
vanishing term $\propto\eta^0$ (this can be {\em explicitly} checked as the
two-body Hamiltonian in the test-mass limit $\eta=0$ is known up to all orders).
As we said above, we expect, as it is the case at 
lower orders, that in $a_5$ the term
$\propto\eta^1$ will dominate.  Let us denote
\begin{equation}
a_{nk} = b_{nk}\,\eta + {\cal O}(\eta^2) \,, \quad n \ge 2 \,.
\end{equation}
The numerical values of the parameters $b_{nk}$ are given in Table 
\ref{table:bnk}.  After studying the various possible patterns
exhibited by  Table~ \ref{table:bnk} 
we decided to focus on the fact 
that the column $b_{n1}$ of Table 
\ref{table:bnk} seems to give a good approximation of the final, total value of 
$b_n$. This would suggest $b_5 \simeq 250$, as possible value.

\begin{table*}[t]
\caption{Numerical values of the coefficients $b_{nk}$.}
\begin{ruledtabular}
\begin{center}
\begin{tabular}{c d d d d d d}
$n$ & b_n & b_{n0} & b_{n1} & b_{n2} & b_{n3} & b_{n4} \\ \hline
2 &  0      &  0.25    &  -0.25   & & & \\
3 &  2      & -0.75    &   1.875  &    0.875 & & \\
4 & 18.6879 &  1.125   &  20.3281 &  -17.125 &  14.3598 & \\
5 &  ?      & -1.05469 & 244.523  & -324.82  & 194.214  & ? \\
\end{tabular}
\end{center}
\end{ruledtabular}
\label{table:bnk}
\end{table*}

Let us now display a more {\em explicit} form (for the parts which are known) of
the 4PN coefficient $a_5=\sum_{k=0}^4 a_{5k}$.  To do this we replace in Eqs.\
\eqref{a50}--\eqref{a54} the 2PN and 3PN coefficients $h_2^k$ and $h_3^k$ by
their general relativistic values.  Out of the 4PN coefficients $h_4^k$ the
leading kinetic term $h_4^5$ it is fully known (as it is given by the expansion
of the free Hamiltonian $\sum_a\sqrt{{\bf p}_a^2+m_a^2}$), for the rest of the
terms only their parts $\propto\eta^0$ are known (they describe the test-mass
limit of the two-body dynamics).  We parametrise our ignorance of the parts
${\cal O}(\eta)$ by introducing some quantities $\chi^k_4$, $k=0,\ldots,4$. Thus
we can write
\begin{subequations}
\begin{eqnarray}
h_4^0 &=& -\frac{1}{16} + \eta \ \chi_4^0 \,,
\\[2ex]
h_4^1 &=& \frac{105}{32} + \eta \ \chi_4^1 \,,
\\[2ex]
h_4^2 &=& \frac{105}{32} + \eta \ \chi_4^2 \,,
\\[2ex]
h_4^3 &=& \frac{13}{8} + \eta \ \chi_4^3 \,,
\\[2ex]
h_4^4 &=& \frac{45}{128} + \eta \ \chi_4^4 \,,
\\[2ex]
h_4^5 &=& \frac{7}{256}(1 - 3\,\eta)(1 - 6\,\eta + 9\,\eta^2 - 3\,\eta^3) \,.
\end{eqnarray}
\end{subequations}
Collecting all this partial  information together one gets
\begin{subequations}
\begin{eqnarray}
a_5 &=& \left(\frac{571}{4}-\frac{197}{64}\pi^2\right) \eta
+ \left(\frac{41}{64}\pi^2-\frac{1885}{96}\right) \eta^2
- \frac{27}{16} \eta^3 + \frac{35}{64} \eta^4 + 2 \eta \sum_{k=0}^4 \chi_4^k
\\[2ex]
&=& 112.3701\,\eta - 13.3127\,\eta^2 - 1.6875\,\eta^3 + 0.5469\,\eta^4
+ 2 \eta \sum_{k=0}^4 \chi_4^k \,.
\end{eqnarray}
\end{subequations}
Note that the expression confirms that the terms $\propto \eta^2,\,\eta^3$ and
$\eta^4$ are sub-dominant.

Based on the above results we guess the range of plausible values for the
parameter $b_5$ to be $[0, 250]$. However, while exploring robustness we would
like to vary $b_5$ beyond this reasonable range subject to the condition that
the potential remains regular.  This condition implies that $b_5 \ge -50$ as
smaller values of $b_5$ introduce poles in the Pad\'e approximated version of
$A(u).$ Note, however, that all the known successive PN approximations suggest
that the $b_n$'s are all positive so that the consideration of negative values
of $b_5$ test robustness against extreme behaviour of the potential.

\subsection{Location of the pole in energy flux $c_P$}
\label{poleSec}

In Ref.\ \cite{DIS1} it was argued that we should expect a (simple) pole in the
flux function as a function of 
\begin{equation}
\label{4.25new}
v \equiv (M\omega)^{1/3} \,,
\end{equation}
where $\omega$ is the orbital frequency, at the location of the light ring.  It
was further shown that factoring out the pole from the post-Newtonian expansion
of the flux before constructing its Pad\'e approximation accelerates convergence
to the fits to the numerical flux.  What happens when we ``flex'' the position
of this pole away from its known test-mass value, or its conjectured
$\eta$-dependent 2PN location?  In the test-mass approximation, that is 
$\eta\rightarrow0$, the location of the pole in the flux function is at 
$v^0_{\rm pole} =1/\sqrt{3}.$ When $\eta$ is different from zero Ref.\  
\cite{DIS1} argued that a good approximation to the location of the pole is 
given by:
\begin{equation}
v_{\rm pole}^\eta = \frac{1}{\sqrt{3}} 
\left( \frac{\dst 1+\frac{1}{3}\eta}{\dst 1-\frac{35}{36}\eta} \right)^{1/2}
\simeq v^0_{\rm pole} \left[1+0.16 \, (4\eta)\right] \,.
\end{equation}
The location of the pole can significantly change the value of the Pad\'e 
approximant of the flux function in the physically relevant region of the 
variable $v$ (see below). For this reason it is important to move the pole away 
from its predicted value and assess how such a shift would affect the 
detectability of the signal. In this work we modify the location of the pole by 
introducing the parameter $c_P$:
\begin{equation}
v_{\rm pole}^{c_P} = \frac{v_{\rm pole}^\eta}{1 + c_P} \,.
\end{equation}
Based on the fact that $v^\eta_{\rm pole}$ differs, when $\eta=1/4$, from the
test mass value $v^0_{\rm pole}$ by $\simeq 16\%$, an {\it a priori} plausible
range of variation of $c_P$ is $\pm 0.2$.  We also explored larger variations of
$c_P$, namely in the range $[-0.5,+0.5]$, which in the comparable mass case
amounts to varying the original 2PN pole from $0.6907$ in the range
$[0.4605,1.3814]$. 
Actually, values of $c_P$ smaller than $-0.2$ seems to have little effect on the
overlaps. [Note that when $c_P$ tends to $-1$ this pushes the pole to
$v_{\rm pole} \rightarrow + \infty.$] If $c_P$ is taken to be greater than about $0.5$ then the location of
the pole in the flux will be at $v<v_{\rm lso}$ so that we will not be able to
compute the phasing of the waves.  This is why we restrict the values of $c_P$
to be smaller than about $0.5.$

\subsection{Unknown third post-Newtonian energy flux $\theta$}
\label{thetaSec}

To estimate the possible range for the unknown parameter $\theta$, let us go
back to Ref.\  \cite{BIJ02} where this arbitrariness is pointed out and
discussed.  The parameter $\theta$ is the linear combination
$\theta=\xi+2\kappa+\zeta$ of the three coefficients $\xi$, $\kappa$ and $\zeta$
associated with three different kinds of terms.  From Eq.~(10.6) of Ref.
\cite{BIJ02} and more explicitly from work in progress \cite{BI02} (which uses
the generalized Hadamard regularization of Ref.\  \cite{BFHad}) it follows that
the value of $\theta$ contains a log term in addition to a term of approximate
value $-1$.  This motivates us to suggest that a variation range by a factor of
order 10 (with both positive and negative signs) is a very generous range for
$\theta$, which is expected to be ``of order unity".

How large can the magnitude of $\theta$ be without introducing any spurious
poles in the P-approximant of the flux?  The answer is that the variation of
$\theta$ is bounded from below at $\theta = -5$ because for $\theta<-5$ there is
a spurious pole.  However, for values $\theta>0,$ even as large as $10^5,$ there
seems to be no irregular behaviour of the P-approximant flux.  Thus, in our test
of robustness we take the minimal range of $\theta$ to be $[-5, 10]$ and we also
explore the values of $\theta>10.$

\subsection{Modification of the two-body Hamiltonian: $z_2$ or $\zeta_2$}
\label{z2Sec}

The 3PN extension of EOB opened the possibility of introducing two free
parameters $z_1$ and $z_2$.  It is clear that taking a non-zero value of $z_1$
goes against the spirit of the EOB resummation, because it takes away a part of
the basic EOB radial potential $A(u)$ to replace it by a modification of the
``centrifugal part'' of the potential.  [See Eqs.\ (\ref{effH2})--(\ref{a4})
above.]  Distributing the 3PN effects between $A(u)$ and the centrifugal
potential is undesirable because it goes against ``resumming'' all effects in
one object:  namely $A(u)$.  Therefore we continue, as in \cite{DJS00}, to fix
$z_1=0$ and this choice simplifies the constraint Eq.\ \eqref{zcon1} to
\begin{equation}
\label{zcon2}
4z_2 + 3z_3 = 24\left(1-\frac{3}{4}\eta\right)\eta \,.
\end{equation}
This leaves us with only one 3PN flexibility parameter linked to this 
possibility and it is convenient to parametrise it by introducing a $\zeta_2$ 
such that,
\begin{subequations}
\begin{eqnarray}
\label{zeta2}
z_2 &=& \frac{3}{4} \, \zeta_2 \, z_3^f \,,
\\
z_3 &=& (1 - \zeta_2) \, z_3^f \,,
\end{eqnarray}
\end{subequations}
where
\begin{equation}
z_3^f \equiv 2\, (4 - 3 \eta) \,.
\end{equation}
Making use of $z_1=0$, and the above parametrisation, the Hamiltonian Eq.\ 
\eqref{effH2} reads
\begin{equation}
\widehat{H}_{\text{eff}}(r,p_r,p_\phi;\zeta_2) = \sqrt{ A(r) \left\{
1 + A(r)D(r;\zeta_2)^{-1}\, p_r^2 + \frac{p_{\phi}^2}{r^2}
+ \frac{z_3^f}{r^2} \left[ \zeta_2 \left(
-\frac{1}{4}p_r^4 +\frac{3}{4} \frac{p_\phi^2 p_r^2}{r^2}\right)  + 
 p_r^4 \right] \right\} } \,, 
\end{equation}
where
\begin{subequations}
\begin{eqnarray}
\label{As} 
A(r) &=& 1 - \frac{2}{r} + \frac{2\eta}{r^3} + \frac{a_4}{r^4} \,, \quad 
a_4 \ = \ \left( \frac{94}{3} - \frac{41}{32}\pi^2 \right) \eta \,,
\\[2ex]
\label{Ds} 
D(r;\zeta_2)^{-1} &=& 1 + \frac{6\eta}{r^2} + 
\frac{4(26-3\eta)\eta-3(4-3\eta)\zeta_2}{2r^3} \,.
\end{eqnarray}
\end{subequations}
The values of $\zeta_2$ equal to 0 and 1 correspond to the simple choices  
$z_2=0$ and  $z_3=0$, respectively. Its variation is of order unity to cover 
this interval and we take its natural range to be $[-2, 2].$

\subsection{Flexibility parameter for non-adiabaticity $f_{\rm NonAdiab}$}
\label{nonadiabSec}

The current version of EOB templates chooses the simplest treatment of
non-adiabatic effects.  In the present study, we would like to look at a
modification of this choice and to this end we introduce the parameter $f_{\rm
NonAdiab}$ in the expression for the angular damping force $\widehat{{\cal
F}}_\phi$ appearing in Eq.~(\ref{eq:dphibydt}).  More precisely, we modify the
current ``minimal'' radiation reaction using:
\begin{equation} 
\label{adia}
\widehat{{\cal F}}_\phi \rightarrow
\widehat{{\cal F}}_\phi \left[ 1 - f_{\rm NonAdiab} \left( 1 + \frac 
{(Au^2)'}{A'} p_\phi^2 \right)
\right].
\end{equation} 
The combination of factors factored by $f_{\rm NonAdiab}$ vanishes in the
adiabatic approximation [see Eq.\ \eqref{pphi}].  The assumption of adiabaticity
is valid for most of the inspiral regime and deviates from it only close to the
last stable orbit.  The modification (\ref{adia}) is a simple way of
parametrising the effect of different choices in the definition of
$\widehat{{\cal F}}_{\phi}$ for orbital motions which start deviating from
adiabaticity.  One generally expects $f_{\rm NonAdiab}$ to be a parameter of
order unity and it suffices {\it a priori} to vary it in the range $[-1,+1].$
While this is the primary goal, we explore a larger range of $f_{\rm NonAdiab}$
in our study of robustness.

\subsection{Non-circular orbits $f_{\rm NonCirc}$}
\label{noncirSec}

The current version of EOB templates also uses a simplest treatment of
non-circular effects, which we would like to re-examine here.  This is
accomplished via another flexibility parameter $f_{\rm NonCirc}$ in the angular
damping force.  We modify the force in a manner similar to the previous case;
that is we use
\begin{equation}
\widehat{{\cal F}}_\phi \rightarrow
\widehat{{\cal F}}_\phi \left[ 1 + f_{\rm NonCirc} \frac{p_r^2}{p_\phi^2 u^2} 
\right].
\end{equation}
As in the previous case, varying $f_{\rm NonCirc}$ in the range $[-1,+1]$ is
expected to be a plausible way of mimicking non-minimal choices of the
definition of $\widehat{{\cal F}}_{\phi}$ for orbital motions which start
deviating from circularity.  However, we do explore a larger range of $f_{\rm
NonCirc}$ in our study of robustness.

\subsection{Transition between plunge and coalescence $f_{\rm transition}$}
\label{transitionSec}

In the EOB approach the equations representing inspiral are continued through
the plunge and eventually matched to the quasi-normal modes of the final black
hole near the light ring.  The exact frequency where this happens cannot be
decided by the formalism and hence one would like to examine the effect on the
waveform by changing the frequency of transition between plunge and coalescence
and subsequent ringing or, equivalently, stopping the plunge at a point
different from the point where the EOB waveform is naturally terminated.  In the
EOB approach the waveform is naturally terminated when the radial coordinate
gets close to the value $r_{\text{light ring}}$ given by a solution to
Eq.\ (\ref{eq:rlightring}).  In the current paper we alter the radial location 
of the light ring by introducing the parameter $f_{\rm transition}$ given by:
\begin{equation}
r_{\text{light ring}} \rightarrow
r_{\text{light ring}} \left( 1 + f_{\rm transition} \right).
\end{equation}

Negative values of the parameter $f_{\rm transition}$ are, rather meaningless
since the EOB approximation is expected to break down for values of the radial
coordinate less than $r_{\text{ light ring}}$.  We therefore allow only positive
values and vary $f_{\rm transition}$ in the range $[0,1].$ Note, however, that
the variation in this parameter is going to seriously affect those systems which
merge in a detector's sensitivity band since a positive value for this parameter
means that we will in effect be discarding power in the final phase of the
signal.  Indeed, in this work, we do not match the plunge waveform to the
quasi-normal mode expected to ensue soon after.  This is because our earlier
work in Ref.\  \cite{DIS3} has shown that these modes do not contribute
significantly to the SNR for those systems whose plunge occurs in the detector's
sensitivity band.

\section{How robust are EOB templates? Visual comparison}
\label{robustnessSec}

In this section we discuss the robustness of EOB waveforms by comparing the
standard fiducial 3PN EOB templates with flexed waveforms constructed by turning
on the flexibility parameters discussed in the previous sections.  We make two
different types of comparisons to gauge the extent to which various unknown
flexibility parameters at third and fourth post-Newtonian orders might affect
the dynamics of the two bodies and the radiation they emit.  Our first
comparison consists of a visual inspection of the behaviour of the relevant
physical quantity when a particular flexibility parameter is varied.  This gives
us an idea of the nature and the extent of the variation involved while testing
robustness.  Our second comparison goes beyond qualitative tests of robustness
by quantitatively measuring the {\it span} of EOB templates.  More precisely, it
consists of the computation of the {\it faithfulness} and {\it effectualness} of
the fiducial EOB template with the flexed waveform and is explored in the
subsequent section.

\subsection{Fourth post-Newtonian dynamics} 

In Sec.\ \ref{b5Sec} we introduced the parameter $b_5$ which encapsulates the
unknown physical effects in the dynamics of the two bodies at orders higher than
the 3PN order.  The most relevant quantity that it affects is the potential
$A(u;b_5)$ which occurs in the effective one-body Hamiltonian and the effective
metric.  Among other things $A(u)$ governs the rate of inspiral of the bodies
and therefore the phase of the waveform.

We begin our visual comparison by plotting the effective potential $A(u)$ at
various PN orders including the 4PN order for two extreme values of $b_5$ 
($b_5=-50$ and $b_5=500$).  Recall that $u=1/r \simeq GM/|{\bf x}_1-{\bf x}_2|$ and 
therefore $u\rightarrow0$ denotes the
region when the two bodies are infinitely separated and $u \simeq 0.5$
denotes the region when the two black holes are ``touching'' each other.  The
sensitivity of ground-based interferometers is best in the frequency range
40--400~Hz.  For a candidate system of total mass $M=20M_\odot$ this frequency
range corresponds to a range for the frequency-related variable $v$,
Eq.\ (\ref{4.25new}), that begins at $v=0.2313$ and terminates beyond $v^{\rm
2PN}_{\rm lso}=0.4457$ \cite{DIS1} or $f_{\rm lso}=286$~Hz.  Equivalently,
this implies a range for the radius-related variable $u$ [using
Eqs.\ (\ref{td12})--(\ref{eq:omegahatVSu}) to connect $v$ and $u$], starting at
$u=0.05335$ or $r=1/u=18.7$. The gray-shaded region corresponds, when the
total mass $M=20 M_\odot$,
to the frequency band $[65,235]$ Hz, centered at 150 Hz, in which the
SNR accumulated for inspirals is more than 80\% of the total
SNR in the entire LIGO band.  For the system $(10M_\odot,10M_\odot)$ the above
frequency band corresponds to $0.2719 \leq v\leq 0.4174,$ equivalently
$0.07372 \le u \le 0.1777$ (or $13.56 \ge r \ge 5.63$).  The dashed vertical
line at $u_{\rm lso} = 0.2065$ near the shaded region corresponds to the
radial coordinate $r_{\rm lso}=4.84$ at which the system reaches the last stable
circular orbit.

It is important to note that the (dashed) vertical line at $u_{\rm lso}$ is
invariant for systems of different masses but the shaded region will change with
the total mass, moving to the right with increasing mass.  The sensitivity of
the instrument is best for those systems for which the LSO is close to ($M\sim
20 M_\odot$) or within ($M\sim 30 M_\odot$) the shaded region.

\begin{figure*}[t]
\centering \includegraphics[angle=-90,width=4in]{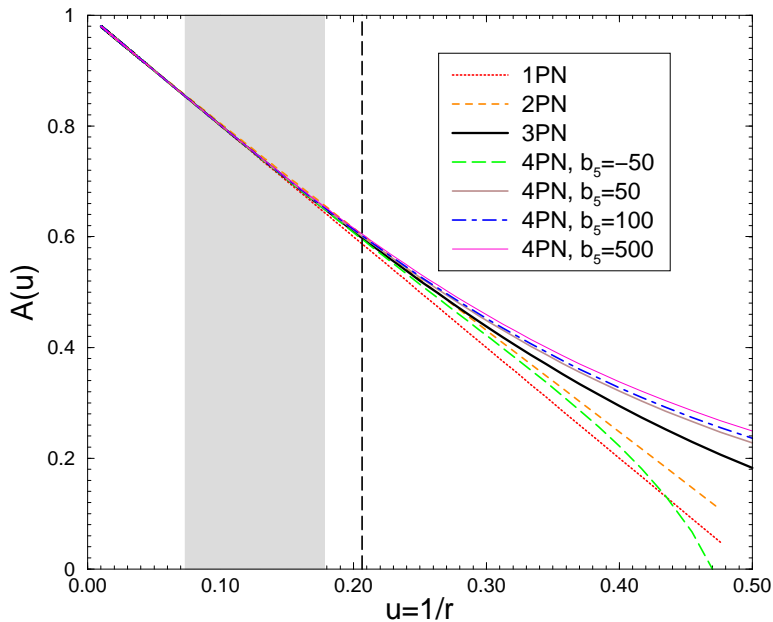}
\caption{The potential $A(u)$ is plotted as a function 
$u=1/r \simeq GM/|{\bf x}_1-{\bf x}_2|$ at various PN orders. By varying the 4PN 
parameter $b_5$ we more than cover the behaviour of both the 2nd and 3rd  
post-Newtonian orders. In all the Figures, for a total mass $M=20 M_\odot$, the 
gray-shaded region corresponds to the frequency band $[65,235]$ Hz, centered at 
150 Hz, in which the signal-to-noise ratio accumulated for inspirals is more 
than 80\% of the total SNR in the entire LIGO band. The corresponding range in 
$u$ is $0.07372 \le u \le 0.1777$ and $r$ is  $13.56 \ge r\ge 5.63$. The 
dashed vertical line at $u_{\rm lso}=0.2065$ 
near the shaded region corresponds to the radial coordinate $r_{\rm lso}=4.84$ 
at which the system reaches the last stable circular orbit.}
\label{potentialFig}
\end{figure*}

{From} Fig.~\ref{potentialFig} we draw three important conclusions:  (a) The
potentials predicted by the two extreme values of $b_5$ used in our study
encompass the variations implied by the second and third post-Newtonian orders.
(b) In the region where the detector is most sensitive to binary black holes the
agreement between the different models is pretty good.  (c) Even consideration
of extremely large positive values of the 4PN parameter $b_5$ has little effect
on the function $A(u)$.  [This is due to the fact that, after Pad\'eing, the
function $A(u)$ has a limit when $b_5\rightarrow\infty$.]  Even variations
beyond reasonable values at $b_5=-50$ and $b_5=500$ lead to an effective
potential that is within the range of variation caused by different
post-Newtonian orders.  These observations already indicate that we should
expect the fiducial EOB template to mimic flexed waveforms reasonably well.

\subsection{Unknown third post-Newtonian energy flux $\theta$}

As mentioned in Sec.\ \ref{3pnDynamicsSec} the gravitational energy flux at
third post-Newtonian order has one undetermined parameter $\theta$. We also
argued in Sec.\ \ref{thetaSec} that the magnitude of $\theta$ should be of order
1.  Since the flux plays a crucial role in the phasing of the waves it is
important to measure the effect of this parameter.  In other words what fraction
of the signal-to-noise ratio will be lost by setting $\theta=0$ in our templates
while in reality $\theta$ is different from zero?  Obviously, the answer depends
on the extent by which $\theta$ is different from zero.  We vary $\theta$ from
$-5$ to $+10$ and plot the Newton-normalized flux as a function of the invariant
velocity parameter $v$. As in Fig.\ \ref{potentialFig} here too the shaded 
region corresponds to a frequency band $[65,235]$ Hz around 150 Hz corresponding 
to the range $0.2719 \leq v \leq 0.4174$ when the total mass $M=20M_\odot$ 
(as assumed in Fig.\ \ref{potentialFig}).  
The dashed vertical line
is of course the ``velocity'' at the last stable orbit $v_{\rm lso} = 0.446$ 
corresponding to two systems of equal mass (independently of the value of the 
total mass).

\begin{figure*}[t]
\centering\includegraphics[angle=-90,width=4in]{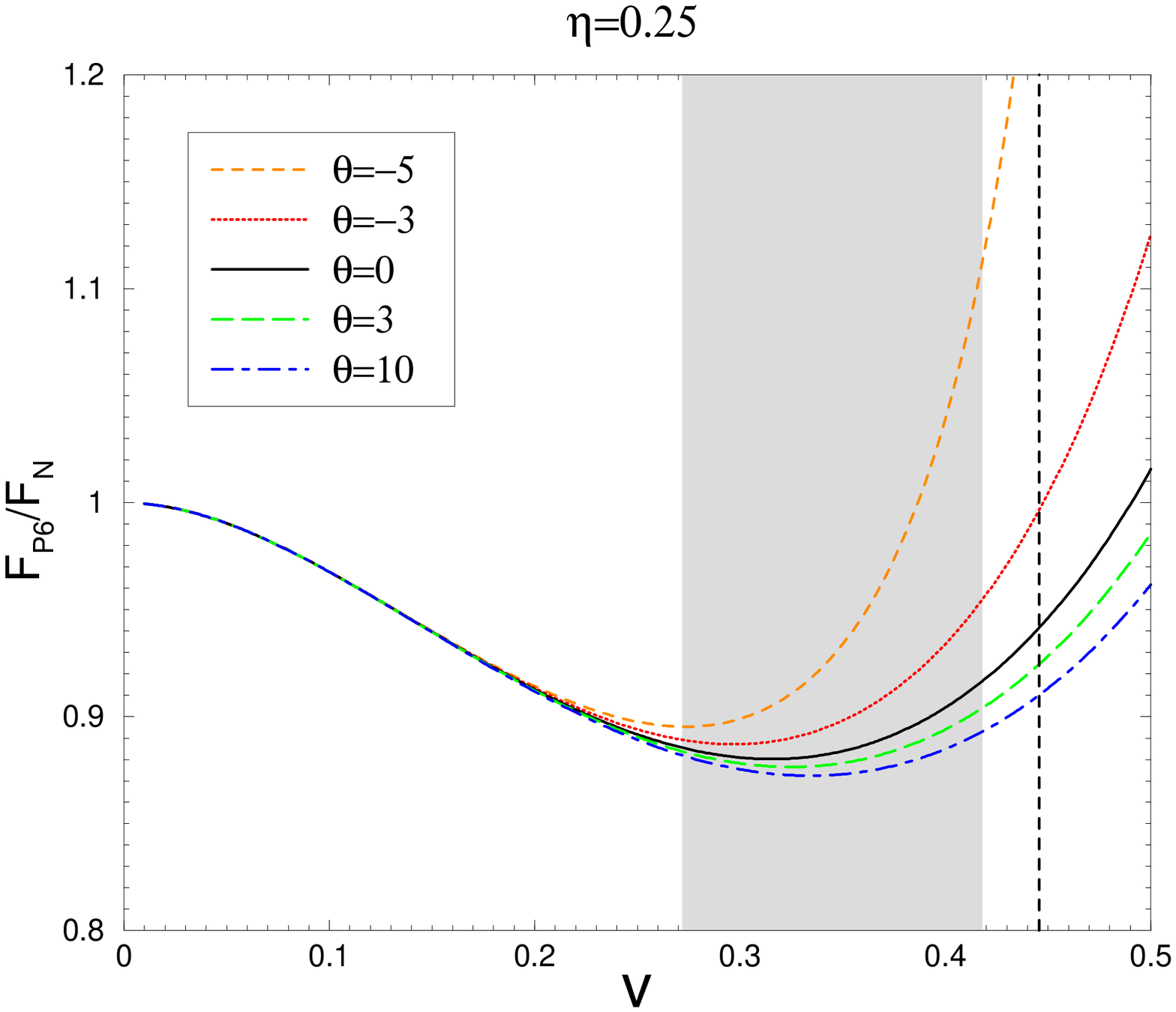}
\caption{Variation in the (Newton-normalized) energy flux emitted by the system 
due to the 3PN parameter $\theta$ being different from zero. Clearly, negative 
values of $\theta$ have a greater effect on the behaviour of the flux as  
compared to the positive values.}
\label{fluxThetaFig}
\end{figure*}

Figure \ref{fluxThetaFig} indicates the extent of variation caused by changing
the value of $\theta.$ Clearly, negative values of $\theta$ have a larger impact
on the flux than positive values.  Indeed, $\theta=-5$ leads to much greater
variation in the flux than even $\theta=+10.$ The main message from
Fig.~\ref{fluxThetaFig} is that by varying $\theta$ over the range $-5\le \theta
\le 10$ in our study of faithfulness and effectualness we would in effect take
into account the possibility that the real gravitational wave flux be rather
different from that assumed in the effective one-body approximation.  (Note,
however, that the differences within the most relevant shaded region are only
$\sim\pm10\%$.)  The variation in $\theta$ we consider is far greater than the
variation of $\pm 2$ in the parameter $\hat \theta$ considered in Ref.
\cite{BCV02}.  Note that $\hat\theta$ is related to our $\theta$ via
\begin{equation}
\hat \theta \equiv \theta + \frac{1987}{1320} \,.
\end{equation}
Indeed our range corresponds to $-3.49 \le \hat\theta \le 11.5$.

\subsection{Flexibility parameter $c_P$}

\begin{figure*}
\centering\includegraphics[angle=-90,width=4in]{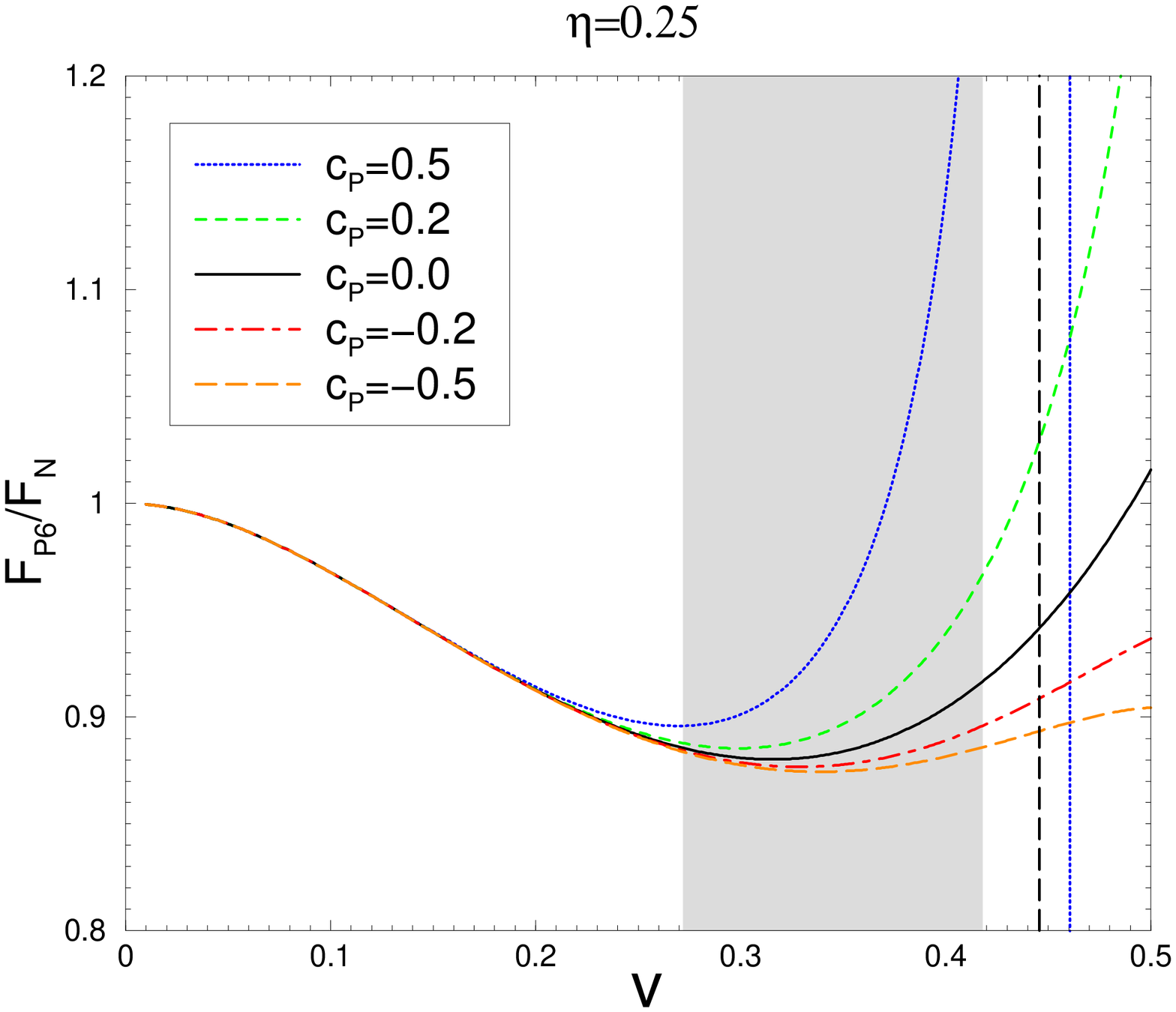}
\caption{The (Newton-normalized) energy flux is plotted for different 
locations of the pole parametrised by $c_P.$ We vary the location 
of the pole by about 50\% on either side of its nominal value 
predicted by the second post-Newtonian binding energy. The range of
$c_P$ is perhaps far greater than what one could expect on physical grounds
and causes a great variation in the flux function. Note that for $c_P=0.5$ the
pole is moved to $v_{\rm pole}=0.4605$ which is near but still beyond the LSO.}
\label{fluxCpFig}
\end{figure*}

In Fig.\ \ref{fluxCpFig} we plot the standard P-approximant flux at 3PN order in
the equal mass case (i.e., $\eta=0.25$, solid line).  In the notation introduced
in Sec.\ \ref{poleSec} this curve corresponds to $c_P=0.$ The effect of changing
the location of the pole is shown by plotting the energy flux for four values of
the parameter $c_P$.  We note that the curves change monotonically as the value
of $c_P$ is changed, moving to the right for negative values of $c_P$ and to the
left for positive values.  We have changed the location of the pole by rougly
50\% on either side of its nominal value.  As mentioned before this amounts to varying the light
ring value of $v$ in the range $0.4605 \le v_{\text{light ring}} \le 1.3814$.
No calculation we are aware of suggests a larger variation in the location of
the pole than considered here.  The curves do show a rather large variety
indicating that the location of the pole is {\it a priori} as important as the
other two parameters discussed before.

\section{The span of the 3PN EOB bank:
Overlap of fiducial template with flexed waveform}
\label{sec:Overlap}

\subsection{Faithfulness and effectualness}
\label{sec:faithfulness}

The ultimate tool for testing the robustness is of course the overlap of  
template waveforms with flexed waveforms.  Given a  fiducial template  
 $T(t; p_k)$ and a  flexed signal   $S(t; p_l, \pi_a)$ their 
overlap ${\cal O}$ is defined as
\begin{equation}
\label{eq:Overlap}
{\cal O}(T,S) \equiv \frac{\langle T\, ,\, S\rangle}
{\sqrt{\langle T\,,\,T \rangle\langle S\,,\,S \rangle}}\,,
\end{equation}
where the scalar product is defined as usual by the Wiener formula
\begin{equation}
\langle X,Y\rangle 
\equiv 2 \int_{0}^{\infty} \frac {df } {S_h(f)}
\left[ \tilde X(f) \tilde Y^*(f) +  \tilde X^*(f) \tilde Y(f) \right].
\label{eq:ScalarProduct}
\end{equation}
Here $\tilde{a}$ denotes the Fourier transform of function $a(t)$, that is, 
$\tilde a(f) = \int_{-\infty}^{\infty} a(t) \exp(-2\pi ift) dt,$ 
$a^*(f)$ denotes complex conjugation of $a(f)$ and $S_h(f)$ is the (one-sided) 
noise spectral density of the detector.  In computing overlaps we use 
the initial LIGO noise spectral density of Ref.\ \cite{DIS3} given by:
\begin{equation}
S_h(f) = 1.44\times 10^{-46} \left [ \left ( \frac{4.64f}{f_k} \right)^{-56}
+ \left ( \frac{f}{f_k} \right )^{-4.52} 
+ 3.25 + 2 \left ( \frac{f}{f_k} \right )^2 \right ]{\rm Hz}^{-1}\,,
\end{equation}
where $f_k=150$~Hz. Since the noise curve rises very steeply at low frequencies 
the lower limit of the integral in Eq.\ (\ref{eq:ScalarProduct}) does not 
have to 
be zero. It suffices to choose a lower limit of 40~Hz so as not to lose more 
than 1\% of the overlap for binaries with total mass $M \simeq 50M_\odot.$

{\it Faithfulness} $\cal F$ is defined as the overlap maximized
only over the
extrinsic parameters of the template, which in our case are simply a reference
time $t_{\rm ref}$ at which the template waveform reaches a certain frequency 
(say 40~Hz)
and the phase $\phi_{\rm ref}$ of the signal at that time:
\begin{equation}
{\cal F} = \max_{t_{\rm ref}, \phi_{\rm ref}} | {\cal O}(T,S) |.
\end{equation}

{\it Effectualness} $\cal E$ is defined as the overlap maximized over not
only the extrinsic parameters but all the intrinsic parameters as well, which
in our case are the two masses $m_1$ and $m_2$ of the binary:
\begin{equation}
{\cal E} = \max_{t_{\rm ref}, \phi_{\rm ref}, m_1, m_2} | {\cal O}(T,S) |.
\end{equation}

\subsection{Matched filtering and signal-to-noise ratio}

In searching for signals of known shape, such as chirping radiation from black
hole binaries, one employs the method of matched filtering.  For signals of
known pattern, matched filtering is, in Gaussian noise background, a
statistically optimum strategy in which a data analyst computes the
cross-correlation of the template waveform $T(t)$ with the detector output
$X(t).$ The analyst will not know before hand when the signal arrives or what
its parameters are.  Therefore, it is necessary to take several copies of the
template corresponding to different parameter values $p_k$ and compute the
correlation of each of those templates with the detector output at different
time-steps $t_0,t_1,\ldots,t_k,\ldots\,$. If detector output contains a
sufficiently strong signal resembling one of the template waveforms then the
cross-correlation will exceed the rms value of the correlation by a large
amount, thereby generating a trigger for the analyst.  It is well-known that the
signal-to-noise ratio $\rho$ of a template $T$ with the detector output that
contains a signal $S$ of known shape is given by
\begin{equation}
\rho\equiv \frac{s}{\overline{n}}=\frac{\vert 
\langle T, S\rangle \vert}{\langle T,T\rangle^{1/2}}
=\vert {\cal O}(T,S)\vert\langle S,S\rangle^{1/2}.
\label{d6}
\end{equation} 
Here $\langle S, S \rangle^{1/2}$ is the signal-to-noise ratio which would be
achievable by matched filtering if the detector output were correlated with the
exact replica of the signal hidden in the noise modulo the amplitude which is
irrelevant.  The above equation tells us that when we do not know the exact
shape of the signal, the signal-to-noise ratio gets degraded and only a fraction
equal to the overlap of the template used in the search with the exact signal
expected to be present in the detector output is what is recoverable.  Thus,
while matched filtering is an excellent technique in detecting signals buried in
noise, dephasing of the template relative to the signal can quickly degrade the
quality of the output.  This can be readily seen from
Eq.~(\ref{eq:ScalarProduct}) where the Fourier {\it amplitudes} of the template
and the signal are multiplied together before being integrated over the
frequency.  These amplitudes coherently add up only when the phase of the
template coincides with that of the signal at all points in the frequency space.
Even a small initial difference in phase can accumulate and kill the integral
since the signals last for a large number of cycles---more than 60 cycles for
$(10 M_\odot$,$10M_\odot)$ black hole binaries and up to 1000 for NS-NS binaries
in the sensitivity band of the LIGO interferometers.  This is the motivation for
building template waveforms that are as close to the true general relativistic
signal as possible.  Note, however, that the weighting of the overlap integral
Eq.~(\ref{eq:ScalarProduct}) by the inverse of the noise spectral density means
that not all cycles in the signal and the template, are equally important.
Ref.~\cite{DIS2} introduced the concept of ``useful cycles'' as a measure of the
effective number of cycles which dominate the overlap integral.  For instance in
the case of a $(10M_\odot,10M_\odot)$ system the number of useful cycles is
$\simeq 7.6$.  This means that it is crucial to model the phasing of the signal
during the $\sim 8$ cycles around the peak of the SNR logarithmic
frequency-distribution i.e.\ around $f_P\simeq165$Hz, but that a less accurate
model of phasing for the other cycles may be acceptable.

\subsection{Span of a template bank}

Stringent as it may sound, matched filtering is not totally restrictive when
dealing with a template bank rather than a single template.  To explain how a
template bank is not as restrictive as a single template we introduce the notion
of {\it span of a template bank.}  In the geometrical language of signal
analysis \cite{BSD96} templates can be thought of as vectors---one vector for
every set of values of the parameters $p_k.$ If the signal depends on $K$
parameters then the set of all vectors run over a $K$-dimensional manifold.
Equation (\ref{eq:ScalarProduct}) serves as a scalar product between different
vectors and induces a natural metric $g_{km}$ on the template-manifold with the
parameters serving as natural coordinates:
\begin{equation}
g_{km} \equiv \langle T_{k}\, , T_{m} \rangle,\ \ T_{k} \equiv \frac{\partial 
T}{\partial p_k}.
\label{eq:metric}
\end{equation}
Though each template is itself a vector in the vector space of all detector
outputs, the set of all templates do not
form a vector space.  Therefore, when dealing with the problem of constructing 
a bank of templates one is really working with only a subspace of the vector space.
Moreover, one does not work with the full template space either but, like in quantum
mechanics, with the set of  {\em rays}, i.e. the set of vectors
modulo their lengths, which can be realized as the set of
{\em normalized} vectors. In other words, we work on
the unit sphere in the initial vector space.
When considering a submanifold on this sphere which
is {\em not}  the intersection of a linear space with the unit
sphere the metric $g_{km}$ gives only a local approximation to the
vector product of the larger space, but it does not endow the
finite-dimensional submanifold with the correct projection of the metric
structure of the larger space.

In our search for gravitational-wave signals we choose a grid of templates on
the template-manifold.  If the templates are an exact replica of the expected
signal then the density of the grid points is so chosen that no signal vector on
the manifold has an overlap with the ``nearest'' grid point smaller than a
certain fraction called the {\it minimal match} MM \cite{SD91,Owen96}, typically
chosen to be either $\text{MM}=0.965$ or possibly $\text{MM}=0.95$:
\begin{equation}
\max_{\text{ Template bank}}\vert {\cal O}(T,S)\vert \ge \text{MM}.
\end{equation}

\begin{table*}[t]
\caption{Robustness of 3PN EOB model with respect to parameters $b_5$, $\theta$, 
$\zeta_2$ and $c_P$. We give the {\it faithfulness} $\cal F$, {\it 
effectualness} $\cal E$, and the total mass $M'$ and symmetric-mass ratio 
$\eta'$ that maximize the overlap while finding effectualness. We vary each 
parameter in the nominal range but also quote extreme values of the parameters 
up to which an overlap of 95\% or greater is obtained. Some parameters, for 
example $b_5$ will be bounded either from below or above since the physical 
quantity it participates in may be irregular for such values thereby affecting 
gravitational-wave phasing.}
\begin{ruledtabular}
\begin {center}
\begin {tabular}{l | c c c | c c c}
\multicolumn{1}{l}{System $\rightarrow $} & 
\multicolumn{3}{c}{$(10M_\odot,10M_\odot)$} & 
\multicolumn{3}{c}{$(15M_\odot,15M_\odot)$} \\[3pt]
\hline\\[-3pt]
parameter $\downarrow$ & ${\cal F}$  & ${\cal E}$ & $(M',\eta')$ & ${\cal F}$ & 
${\cal E}$ & $(M',\eta')$ \\[7pt]
\hline\\[-8pt]
$b_5= -50$   & 0.9754        & 0.9944  & (20.25, 0.2479)\ \ \ \ & 0.9728    & 
0.9953 & (30.41, 0.2494) \ \ \ \ \\
$b_5= +50$   & 0.9431        & 0.9850  & (19.76, 0.2499)\ \ \ \ & 0.9448    & 
0.9943 & (29.25, 0.2499) \ \ \ \ \\
$b_5= +190$   & 0.8566       & 0.9663  & (19.61, 0.2500)\ \ \ \ & 0.8855   & 
0.9829 & (28.92, 0.2500) \ \ \ \ \\
$b_5= +250$   & 0.8359        & 0.9633  & (19.63, 0.2485)\ \ \ \ & 0.8476    & 
0.9766 & (28.54, 0.2500) \ \ \ \ \\
$b_5=+300$   & 0.8014        & 0.9558  & (19.48, 0.2500)\ \ \ \ & 0.8363    & 
0.9753 & (28.46, 0.2500) \ \ \ \ \\
$b_5=+500$   & 0.7849        & 0.9498  & (19.46, 0.2499)\ \ \ \ & 0.8181    & 
0.9694 & (28.46, 0.2499) \ \ \ \ \\[10pt]
$c_P=-0.2$   & 0.9988        & 0.9999  & (19.97, 0.2500)\ \ \ \ & 0.9993    & 
0.9998 & (29.96, 0.2500) \ \ \ \ \\
$c_P=+0.2$   & 0.9941        & 0.9989  & (20.17, 0.2469)\ \ \ \ & 0.9975    & 
0.9999 & (30.31, 0.2469) \ \ \ \ \\[10pt]
$\theta=-5 $ & 0.9358        & 0.9949  & (20.27, 0.2500)\ \ \ \ & 0.9642    & 
0.9971 & (30.77, 0.2497) \ \ \ \ \\
$\theta=+10$ & 0.9946        & 0.9982  & (19.97, 0.2500)\ \ \ \ & 0.9964    & 
0.9999 & (29.86, 0.2498) \ \ \ \ \\[10pt]
$\zeta_2=-2$     & 0.9999        & 1.0000  & (20.01, 0.2497)\ \ \ \ & 0.9998    
& 0.9998 & (30.02, 0.2493) \ \ \ \ \\
$\zeta_2=+2$     & 0.9999        & 1.0000  & (20.01, 0.2500)\ \ \ \ & 0.9998    
& 0.9999 & (30.08, 0.2490) \ \ \ \ \\[10pt]
\end {tabular}
\end {center}
\end{ruledtabular}
\label{table:robustness1}
\end {table*}

The above inequality will be satisfied not only for signal vectors on the
template manifold but also vectors that are {\it off} the manifold but close to
it.  In other words, the template bank obtains minimal match for all signals
located in an infinite dimensional ``slab'' around the $K$-dimensional template
manifold but sufficiently close to it.  This ``slab'' defines the {\it span} of
the considered template bank, say ${\cal S}[T,{\rm MM}]$.  Here, for simplicity
we introduce one flexibility parameter at a time and explore successively the
``slab'' along the directions defined by each extra parameter.  It therefore
suffices to consider only those signals that live in a $(K+1)$-dimensional space
around the template manifold, which particular $K+1$, depending on the
flexibility parameter in question.  The {\it span} ${\cal S}[T,{{\rm MM}}]$ of a
template bank along a given flexibility direction is then defined as the maximum
domain in the corresponding $(K+1)$-dimensional space within which the template
bank $T$ obtains a given minimal match {MM}.  In this work we estimate this
domain by computing the range of the flexibility parameters within which the
minimal match is achieved between the fiducial template and flexed waveforms.

When the template is not a true representation of the signal, the signal vectors
run over a manifold that does not exactly coincide with the template manifold.
What is required for signal detection is that the span of the template bank
includes the signal-manifold.  Of course, if the minimal match is sufficiently
small then any template bank would span the signal-manifold.  Successful signal
detection, without undue loss of signals, requires the signal-manifold to be a
submanifold of ${\cal S}[T, 0.95]$ (i.e., 95\% minimal match) or, better, of ${\cal
S}[T, 0.965]$ (i.e., 96.5\% minimal match).

Finally, let us note that while the span is defined with respect to a continuum of
template bank in reality we will have to be content with a finite lattice of templates.
Therefore, it is not guaranteed that the (maximum of the) overlap of a finite template 
bank with an arbitrary flexed signal within the span ${\cal S}[T,\ 0.965]$ 
will be greater than 0.965.  If the template lattice is chosen such that 
the minimal match is at least 0.965 then the maximum overlap reached within
the span ${\cal S}[T,\ 0.965]$ of the template bank might be reduced to 
$0.965^2 \simeq 0.93.$ So the actual loss in the event rate might be 
$1-(0.965^2)^3\simeq 20\%.$

\subsection {Span of third post-Newtonian EOB template bank}

The main conclusion of this study is that the standard 3PN EOB template bank 
without any additional parameter,
{\it spans} the extensions separately implied by the {\it seven flexibility
parameters} that account for the {\it  unmodelled effects
in the EOB formalism affecting both the dynamics
and radiation flux.}  This is demonstrated in Tables \ref{table:robustness1}
and \ref{table:robustness2} where we have considered two archetypal binaries
expected to be observed by initial interferometers---the first consisting of a
pair of $10M_\odot$ black holes and the second consisting of two $15 M_\odot$
black holes.  In Table \ref{table:robustness1} we have explored the faithfulness
and effectualness of  the 3PN EOB template bank (our fiducial template)
with respect to the four important
flexibility parameters $b_5,$ $c_P,$ $\theta$, and $\zeta_2$ and in Table
\ref{table:robustness2} the same, but for the less important flexibility
parameters $f_{\rm NonAdiab},$ $f_{\rm NonCirc},$ and $f_{\rm transition}.$

First, we discuss the results obtained by varying the parameters over the range
in which they are expected to lie.  Clearly, the 4PN parameter $b_5$ has the
strongest influence followed by the parameters $c_P, \theta$ and $\zeta_2.$
Indeed, the faithfulness is not always larger than the fiducial minimal match of
0.965 when these parameters take values over the range in which they are
expected to vary.  However, for $(15 M_\odot,15 M_\odot)$ systems the {\it
effectualness} does easily meet the usual requirement ${\cal E}> 0.965$ for {\it
all values} of $b_5$ [including $b_5>500$, in view of the fact, visible on Fig.\
\ref{potentialFig}, that $A(u;b_5)$ monotonically reaches a smooth limit as
$b_5\rightarrow+\infty$].  In the case of $(10M_\odot,10M_\odot)$ systems the
situation is a bit more involved:  (1) when $b_5\lesssim 200$, the usual
requirements on effectualness is met, but (2) when $b_5>200$, the effectualness
drops slightly below $0.965$.  However, for the ``plausible'' value $b_5=250$
(see Sec.~\ref{sec:unknown physics} above) the overlap is still larger than
$0.963$, and even for $b_5=500$ (and probably for any larger $b_5$ for the
reason mentioned above) the overlap is still as large as $0.95$.  Note also
that, in many cases, faithfulness is itself larger than the minimal match and
effectualness is close to 1.  

In view of this special sensitivity to $b_5$, we
explored in detail the $b_5$-dependence of overlaps and found a simple
modification of the standard 3PN EOB templates that allows for meeting the
desired requirement ${\cal E}>0.965$ for {\it all values} of $b_5$.  If one
constructs a fiducial template bank by using as EOB potential the
``$b_5$-flexed'' function $A_{50}(u)\equiv A_{\text{4PN}}(u;b_5=50)$ [instead of
$A_{\text{3PN}}(u)$], we have found that it leads to effectualness larger than
0.965 in all cases [and in particular for the $(10M_\odot,10M_\odot)$ system and
$-50\leq b_5\leq 250$].  Furthermore, as illustrated in Table \ref{table:max
range}, the span of this new fiducial template bank now extends over all the
values of $b_5$:  $-50\leq b_5\leq 2000$.  In the case of the less important
parameters $f_{\rm NonAdiab},$ $f_{\rm NonCirc},$ and $f_{\rm transition},$ we
observe that the faithfulness is itself larger than our minimal match except
when $f_{\rm transition}=1$ for the $(15M_\odot,15M_\odot)$ system.

As mentioned in Sec.~\ref{sec:unknown physics} we have also explored robustness
beyond the range in which the flexibility parameters are expected to lie and yet
achieve the required effectualness.  In Table \ref{table:max range} we summarize
the range over which the different parameters are expected to vary together with
the range over which they can be varied yet maintaining an effectualness of
0.965.  This table shows that the span of the 3PN EOB bank of templates (or for
that matter, the $b_5=50$-flexed 3PN one) in the other flexibility directions
extend well beyond the expected plausible ranges.

\subsection{How could the match be good when the flux functions look so very 
different?}

In Sec.~\ref{robustnessSec} we noted that the behaviour of the energy flux
$F_{P_n}(v)/F_N(v)$ could be significantly different from their usual behaviour
when the flexibility parameters are set to extreme values in their expected
range.  When the flux is so different how is it still possible to achieve good
effectualness?

The answer lies in several aspects of the problem:  First, one should note that,
after factorisation of the crucial ``quadrupolar flux'' $\propto v^{10}$, the
changes in the Newton-normalized flux are less than $10\%$.  Second, one should
remember that for the massive binaries considered here, the number of ``useful''
gravitational-wave cycles \cite{DIS2} corresponding roughly to shaded regions in
Figs.\ \ref{fluxThetaFig} and \ref{fluxCpFig} is quite moderate ($\lesssim8$).  
Third, one should note that one of the crucial things affected by the flux
is the total chirp time, or the duration, of the waveform.  For instance, if the
flux increases more rapidly when one of the flexibility parameters is nonzero,
as in the case of $\theta=-5,$ then the system loses energy more rapidly and
therefore the waveform lasts shorter.  However, this shortening of the waveform
can also be achieved by making the binary heavier (or lengthened by making
asymmetric binaries of the same total mass or simply lighter binaries).  Recall
that the effectualness is obtained by maximizing the overlap over both the
extrinsic and intrinsic parameters.  Thus, in the process of maximization one
can absorb a change in time-scale by choosing a binary of different total mass
and mass ratio.
This explanation is borne out by a comparison of the trend of the curves in
Figs.\ \ref{potentialFig}, \ref{fluxThetaFig} and \ref{fluxCpFig} with the
corresponding rows in Table \ref{table:robustness1}.  For instance,
effectualness for $\theta>0$ (smaller flux at a given $v$ than when $\theta=0$)
requires a system of total mass lighter than the original system while for
$c_P>0$ (greater flux at a given $v$ than when $c_P=0$) one requires a heavier
system than the original one.

A final comment:  The insensitivity of the effectualness to the location of the
pole can be interpreted to mean that the factorisation by the pole is not as
crucial an element of the gravitational-wave flux resummation as perceived in
\cite{DIS1}.
This suggests that it would be interesting to study the performance of templates 
which do not use such a factorisation of the flux function. We leave this study 
to future work.

\begin{table*}
\caption{Robustness of 3PN EOB model with respect to parameters
${f}_{\rm NonAdiab}$, ${f}_{\rm NonCirc}$, and ${f}_{\rm transition}$. We give 
the {\it faithfulness} $\cal F$, {\it effectualness} $\cal E$, and the total 
mass $M'$ and symmetric-mass ratio $\eta'$ that maximize the overlap while 
finding effectualness. We vary each parameter in the nominal range but also 
quote extreme values of the parameters up to which an overlap of 95\% or greater 
is obtained.}
\begin{ruledtabular}
\begin {center}
\begin {tabular}{l | c c c | c c c}
\multicolumn{1}{l}{System $ \rightarrow$}  & 
\multicolumn{3}{c}{$(10M_\odot,10M_\odot)$} & 
\multicolumn{3}{c}{$(15M_\odot,15M_\odot)$ } \\[3pt]
\hline\\[-3pt]
Parameter $ \downarrow$ & ${\cal F}$  & ${\cal E}$ & $(M',\eta')$ & ${\cal F}$ & 
${\cal E}$ & $(M',\eta')$ \\[7pt]
\hline\\[-8pt]
${f}_{\rm NonAdiab}=-1.0$  & 0.9976 & 1.0000 & (20.04, 0.2491) \ \ \ \ & 0.9967 
& 1.0000 & (30.11, 0.2486) \\
${f}_{\rm NonAdiab}=+1.0$  & 0.9979 & 0.9999 & (20.00, 0.2500) \ \ \ \ & 0.9964 
& 0.9999 & (30.01, 0.2496) \\[10pt]
${f}_{\rm NonCirc}=-1.0$   & 0.9998 & 0.9998 & (20.09, 0.2473) \ \ \ \ & 0.9997 
& 0.9999 & (30.02, 0.2500) \\
${f}_{\rm NonCirc}=+1.0$   & 0.9998 & 0.9999 & (20.10, 0.2472) \ \ \ \ & 0.9996 
& 1.0000 & (30.04, 0.2500) \\[10pt]
${f}_{\rm transition}=0.5$ & 1.0000 & 1.0000 & (20.00, 0.2500) \ \ \ \ & 0.9967 
& 0.9994 & (30.03, 0.2498) \\
${f}_{\rm transition}=1.0$ & 0.9877 & 0.9878 & (20.01, 0.2500) \ \ \ \ & 0.9531 
& 0.9606 & (30.26, 0.2498) \\
\end {tabular}
\end {center}
\end{ruledtabular}
\label{table:robustness2}
\end {table*}

\subsection{Systematic versus statistical errors in the estimation of 
parameters}

Together with the value of the effectualness, the tables also show the template
parameters that obtain the maximized overlaps.  A quick inspection reveals that
the symmetric mass ratio $\eta$ of the template that obtained the maximum match
is either equal to the actual value of 1/4 or when different from 1/4 the
fractional difference is less than 0.1\%.  This is probably explained by the
following:  We study templates as functions of $m_1$ and $m_2$.  But the
function $T(m_1,m_2)$ is invariant under the permutation $m_1\leftrightarrow 
m_2$ and therefore the overlap ${\cal O}(m_1,m_2)=\langle T(m_1,m_2),S\rangle$
always reaches an extremum (along the lines $m_1+m_2=\text{constant}$) at
$m_1=m_2$, i.e.\ at $\eta=1/4$.  If these extrema are all maxima (as a
function of the ratio $m_1/m_2$, for a fixed value of $m_1 + m_2$), the real
maximum of the overlap must lie somewhere along the ``ridge'' $m_1=m_2$.  Note,
however, that there might as well be domains of parameter space where the
overlap ${\cal O}(m_1,m_2)$ reaches a minimum along the ridge at $m_1=m_2$ (at
fixed mass scale $m_1+m_2=\text{constant}$).  The total mass is different from
the true total mass at worst by about 1.5\%.  These percentages, of course, do
not give us a measure of the accuracy of estimation of the parameters, rather
they tell us the extent of {\it bias} induced in the estimation.  Since our
template bank contains waveforms that are not exactly the same as the ``true''
signals, the parameters that maximize the overlap are different from the real
values meaning there is a systematic error in the estimation of parameters and
the percentages we have quoted are upper limits on the {\it systematics.}

The {\it statistical} errors in the estimation of the intrinsic parameters
$m_1,m_2$ are determined by the shape of the level contours of the overlap
function between the template $T(m_1,m_2)$ and the signal, that we assume here
to be part of the template bank:  $S=T(m^0_1,m^0_2)$.  For high signal-to-noise
ratio this shape is an ellipse (the error ellipse) which is determined by the
information matrix, i.e.\  the metric Eq.~(\ref{eq:metric}).  We have
confirmed that even for the massive binary systems that we consider, these error
contours are qualitatively well described by the analytical results of
\cite{PoissonAndWill}, i.e.\  that, when represented in the $({\cal
M},\eta)$-plane, where ${\cal M}=\eta^{3/5} M$ is the {\it chirp} mass, they are
highly elongated ellipses with major axis roughly along the $\eta$-direction and
minor axis along the ${\cal M}$-direction.  When represented in the
$(m_1,m_2)$-plane these error contours have the shape of thin crescents
orthogonal to the diagonal $m_1=m_2$.  A bad consequence of this fact is that
there can be a large statistical error in the determination of $\eta$
\cite{PoissonAndWill}, and therefore correspondingly large statistical errors in
the determination of the individual masses.  Only the combination ${\cal
M}=\eta^{3/5} M$ might be reasonably free of statistical errors.  A more
detailed analysis is needed to assess the total errors combining systematic and
statistical effects.

We note in passing that a useful consequence of the highly elongated structure
of the overlap contours is to allow a fast first-cut data analysis based on the
lower-dimensional template bank defined by fixing $\eta$, {\it e.g.}  to $1/4$,
and varying only $M=m_1+m_2$.  The choice $\eta=1/4$ would be sufficient to
cover systems in a large domain of mass space around the diagonal $m_1=m_2$.  It
can then be complemented by considering a few other simple values of $\eta$.
Each such lower-dimensional (approximate) template bank (corresponding to some
value of $\eta$) finally depends on only one universal function of one variable,
the scaled phasing function $\phi_{\eta}(\hat{t})$, obtained by integrating,
once for all, the EOB equations of motion expressed in terms of scaled variables
$\hat{t}=t/M,\,r \simeq |{\bf x}_1-{\bf x}_2|/GM$, for a particular $\eta,$ 
extending to EOB the idea of {\it mother templates} for the post-Newtonian 
model \cite{BSS00}.  The bank
of templates is then built from the shifted and scaled function
$\phi_\eta((t-t_0)/M)$.  This fact should simplify further the filter bank
construction similar to the case of Newtonian signals which were expressible in
terms of the universal function $\phi_N(\hat{t}) \propto -(-\hat{t})^{5/8}$.
Indeed, the fact that this must be so is implicit in the nature of the template
bank constructed by several authors (cf. second reference in \cite{Owen96} and Ref.\cite{BSS94}) 
who find that a small range of $\eta$ is needed for a large range of the total mass.

\begin{table*}
\caption{The natural range for the flexibility parameters expected on physical 
grounds is shown together with the actual span for which the effectualness is 
greater than 0.965 for equal mass binaries of total mass 20$M_\odot$ and  
$30M_\odot$. As discussed in the text the span in the $b_5$-direction refers to
a bank of templates constructed with slightly modified EOB potential: 
$A_{50}(u)$.}
\begin{ruledtabular}
\begin{center}
\begin{tabular}{l | c c c c c c c}
\\[-3pt]
Range/Span & $b_5$  & $c_P$ & $\theta$ & $\zeta_2$ & $f_{\rm NonAdiab}$ & 
$f_{\rm NonCirc}$ & $f_{\rm transition}$
\\[5pt]
\hline\\[-5pt]
Expected Range    & $[0, 250]$  & $[-0.2, 0.2]$  & $[-5, 10]$  & $[-2, 2]$  & 
$[-1, 1]$  & $[-1, 1]$  & $[0, 0.5]$ \\[5pt]
Span ${\cal S}(T,0.965)$ & $[-50, 2000]^{\*}$  & $[-0.5, 0.5]$  &
$[-5, >\!\!1000]$  & $[-5, 100]$  & $[-20, 15]$  & $[-50, 50]$  & $[0, 1]$ 
\\[3pt]
\end{tabular}
\end{center}
\end{ruledtabular}
\label{table:max range}
\end{table*}

\section{Conclusions}

In this study we have explored the robustness of 3PN EOB templates.  We
introduced seven flexibility parameters that affect the two-body dynamics and
radiation emission and varied each of them separately over a range that can be
reasonably considered to be large enough to encompass unknown and unmodelled PN
effects.  The parameters introduced are:  (a) a 4PN parameter $b_5$ that alters
the two-body effective metric and the EOB potential $A(u)$.  
We conducted a special study of the structure of 4PN
contributions to the Hamiltonian to estimate the plausible range of the parameter $b_5$
measuring them ($-50\le{b_5}\le250$).  (b) the unknown 3PN parameter $\theta$ affecting the nature of
the energy flux emitted by the system ($-5\le\theta\le10$).  (c) a parameter
$\zeta_2$ that changes ${\cal O}({\bf p}^2p_r^2)$ terms in the two-body
Hamiltonian ($-2\le{\zeta_2}\le+2$).  (d) location of the pole in the energy
flux controlled by a parameter $c_P$.  (e) and three parameters $f_{\rm
NonAdiab}$, $f_{\rm NonCirc}$, and $f_{\rm transition}$, that are varied so as
to inflict at least a factor two change in the modelling of non-adiabatic
effects, non-circular effects and the transition from inspiral to plunge,
respectively.

We then compared the faithfulness and effectualness of standard fiducial EOB
templates (that is, EOB templates in which all the above parameters are set
equal to zero) with the flexed signals obtained by switching on the flexibility
parameters one at a time.  Based on the study conducted in this work we find
that the third post-Newtonian EOB templates lead to effectualness larger than
96.3\% in all cases (when $b_5\leq 250$).  For the $(10M_\odot,10M_\odot)$
systems, and $b_5$ larger than 190 the effectualness drops below the usual
requirement ${\cal E}>0.965$, though it remains very close to it, being larger
than 0.96.  (Even when $b_5$ gets very large the effectualness never drops below
0.95.)  

There are two ways of improving this situation linked to the special
sensitivity to $b_5$.  One way is to augument the standard 3PN bank of templates
[based on $A_{\text{3PN}}(u)$] by (when it is needed) a second bank of
templates, based on $A_{100}(u)\equiv A_{\text{4PN}}(u;b_5=100)$.  We have
checked that this ``doubled'' bank of templates allows one to span all values of
$b_5$ with overlaps better than 0.985. A second way (which minimizes
the total number of templates needed) is to work in all cases with only one
specific $b_5$-flexed bank of templates [namely the one based on the
``intermediate'' EOB potential $A_{50}(u)\equiv A_{\text{4PN}}(u;b_5=50)$].  The
remarkable agreement between numerical and analytical descriptions of circular
orbits near the LSO \cite{DGG02,B02a}, suggests that it might be possible soon
to use numerical simulations to map in detail the EOB potential $A(u)$ near the
LSO.  Hopefully this might lead to a numerical estimate of the value of $b_5$
thereby sharpening the preferred choice of bank of templates.

There is a caveat in the current evaluation which we must bear in mind namely
that we vary the flexibility parameters only one at a time. It is possible that the
actual physical signal has more than one of the flexibility parameters non-zero.
In that case our fiducial templates might not be able to achieve the desired span.
This is because the `shifts' in the fiducial template parameters needed to separately
correct for the various non-zero flexibility parameters might be in different
`directions'.  An attempt to define a rather phenomenological signal was
the main focus of Ref. \cite{BCV02}. In this paper, we have focussed on a different 
(minimalistic) attitude and have not looked into these  matters.
We intend to address this issue in a future work.

To conclude:  the 3PN EOB templates (possibly suitably $b_5$-flexed) are good
models to use for black hole binary searches in the interferometer
gravitational-wave data because their ``span'' in signal space seems large
enough to encompass most of the plausible modifications one can think of making
in the current EOB framework.  Moreover, as we emphasized, the highly elongated
shape of overlap contours in the $(M,\eta)$-plane suggests the interesting
possibility to drastically reduce the number of EOB templates by using a small
number of universal phasing functions $\phi_{{\eta}_{i}}(\hat{t})$
with a small discrete set of values of $\eta$ prominently including $\eta=1/4$.

\acknowledgments

We are grateful to Luc Blanchet and Alessandra Buonanno for a critical reading
of the manuscript and comments.
BSS thanks the Max-Planck Institut f\"ur Gravitationsphysik where most of the
work reported here was done while the author was on a sabbatical.  BRI, PJ and
BSS thank the Institut des Hautes Etudes Scientifiques for hospitality during
the course of this work.  This research was in part funded by PPARC grant
PPA/G/O/1999/00214 (to BSS) and by the Polish KBN Grant No.\ 5 P03B 034 20 (to
PJ).


\begin{thebibliography}{999}

\bibitem{glpps}
L.\ P.\ Grishchuk, V.\ M.\ Lipunov, K.\ A.\ Postnov, M.\ E.\ Prokhorov,
and B.\ S.\ Sathyaprakash,
Phys.\ Usp.\ {\bf 44}, 1 (2001); Usp.\ Fiz.\ Nauk {\bf 171}, 3 (2001).

\bibitem{BCV02}
A.\ Buonanno, Y.\ Chen, and M.\ Vallisneri,
{\it Detection template families for gravitational waves from the final stages 
of binary-black-hole inspirals. I. Nonspinning case}, Phys.\ Rev. \ D (2002)
(To Appear);
gr-qc/0205122.

\bibitem{BD99}
A.\ Buonanno and T.\ Damour,
Phys.\ Rev.\ D {\bf 59}, 084006 (1999).

\bibitem{BD00}
A.\ Buonanno and T.\ Damour,
Phys.\ Rev.\ D {\bf 62}, 064015 (2000). 

\bibitem{DJS00}
T.\ Damour, P.\ Jaranowski, and G.\ Sch\"afer,
Phys.\ Rev.\ D {\bf 62}, 084011 (2000).

\bibitem{TD02}
T.\ Damour,
Phys.\ Rev.\ D {\bf 64}, 124013 (2002).

\bibitem{DIS3}
T.\ Damour, B.\ R.\ Iyer, and B.\ S.\ Sathyaprakash, 
Phys.\ Rev.\ D {\bf 63}, 044023 (2001).

\bibitem{BGG02}  E. Gourgoulhon, P. Grandcl\'ement, S. Bonazzola,
Phys.\ Rev.\  D {\bf 65},  044020 (2002);
P. Grandcl\'ement, E. Gourgoulhon, S. Bonazzola, 
Phys.\ Rev.\  D {\bf 65},  044021 (2002).

\bibitem{DGG02}
T.\ Damour, E.\ Gourgoulhon, and P.\ Grandcl\'ement, 
Phys.\ Rev.\ D {\bf 66}, 024007 (2002).

\bibitem {B02a}
L.\ Blanchet,
Phys.\ Rev.\ D {\bf 65}, 124009 (2002).

\bibitem{DIS1}
T.\ Damour, B.\ R.\ Iyer, and B.\ S.\ Sathyaprakash,
Phys.\ Rev.\ D {\bf 57}, 885 (1998).

\bibitem{cutleretal93a}
C.\ Cutler {\it et al.},
Phys.\ Rev.\ Lett.\ {\bf 70}, 2984 (1993).

\bibitem{3PNADM1}
P.\ Jaranowski and G.\ Sch\"afer, 
Phys.\ Rev.\ D {\bf 57}, 7274 (1998); {\bf 60}, 124003 (1999).

\bibitem{3PNADM2}
T.\ Damour, P.\ Jaranowski, and G.\ Sch\"afer, 
Phys.\ Rev.\ D {\bf 62}, 044024 (2000);
{\bf 62}, 021501(R) (2000) [Erratum: {\bf 63}, 029903(E) (2001)];
{\bf 63}, 044021 (2001).

\bibitem{3PNBF}
L.\ Blanchet and G.\ Faye, 
Phys.\ Lett.\ A {\bf 271}, 58 (2000);
Phys.\ Rev.\  D {\bf 63}, 062005 (2000). 

\bibitem{BFHad}
L.\ Blanchet and G.\ Faye, 
J.\ Math.\ Phys.\ {\bf 41}, 7675 (2000); {\bf 42}, 4391 (2001).

\bibitem{ABF01}
V.\ C.\ de~Andrade, L.\ Blanchet, and G.\ Faye,
Class.\ Quantum Grav.\ {\bf 18}, 753 (2001).

\bibitem{BIcm}
L. Blanchet  and  B. R. Iyer,  
Class. Quantum Grav. {\bf 20}, 755 (2003)

\bibitem {DJS01}
T.\ Damour, P.\ Jaranowski, and G.\ Sch\"afer,
Phys.\ Lett.\ B {\bf 513}, 147 (2001).

\bibitem{MPM} 
L.\ Blanchet and T.\ Damour,
Phil.\ Trans.\ R.\ Soc.\ London {\bf A 320}, 379 (1986);
L.\ Blanchet, Proc.\ R.\ Soc.\ Lond.\ {\bf A 409}, 383 (1987);
L.\ Blanchet and T.\ Damour, Phys.\ Rev.\ D {\bf 37}, 1410 (1988);
Ann.\ Inst.\ H.\ Poincar\'e (Phys.\ Th\'eorique) {\bf 50}, 377 (1989);
T.\ Damour and B.\ R.\ Iyer,
Ann.\ Inst.\ H.\ Poincar\'e (Phys.\ Th\'eorique) {\bf 54}, 115 (1991);
Phys.\ Rev.\ D {\bf 43}, 3259 (1991);
L.\ Blanchet and T.\ Damour, Phys.\ Rev.\ D {\bf 46}, 4304 (1992);
L.\ Blanchet, Class.\ Quantum Grav.\ {\bf 15}, 1971 (1998).

\bibitem{2PN}
L.\ Blanchet, T.\ Damour, B.\ R.\ Iyer, C.\ M.\ Will, and A.\ G.\ Wiseman,
Phys.\ Rev.\ Lett.\ {\bf 74}, 3515 (1995);
L.\ Blanchet, T.\ Damour, and B.\ R.\ Iyer, Phys. Rev. D {\bf 51}, 5360 (1995);
C.\ M.\ Will and A.\ G.\ Wiseman, Phys.\ Rev.\ D {\bf 54}, 4813 (1996);
L.\ Blanchet,  B.\ R.\ Iyer, C.\ M.\ Will, and A.\ G.\ Wiseman,
Class.\ Quantum Grav.\ {\bf 13}, 575, (1996);
L.\ Blanchet, Phys.\ Rev.\ D {\bf 54}, 1417 (1996).

\bibitem{btail}
L.\ Blanchet,
Class.\ Quantum Grav.\ {\bf 15}, 113 (1998).

\bibitem{BFIJ02}
L.\ Blanchet, G.\ Faye, B.\ R.\ Iyer, and B.\ Joguet,
Phys.\ Rev.\ D {\bf 65}, 061501(R) (2002).

\bibitem{BIJ02}
L.\ Blanchet, B.\ R.\ Iyer, and B.\ Joguet,
Phys.\ Rev.\ D {\bf 65}, 064005 (2002).

\bibitem{DIS4}
T.\ Damour, B.\ R.\ Iyer, and B.\ S.\ Sathyaprakash,
Phys.\ Rev.\ D {\bf 66}, 027502 (2002).

\bibitem{LBLR}
L.\ Blanchet, in Proceedings  of the 25th Johns Hopkins Workshop, 
Eds. I. Ciufolini and L. Lusanna; gr-qc/0207037.

\bibitem{BI02}
L.\ Blanchet and B.\ R.\ Iyer,
{\it work in progress}.


\bibitem{BSD96}
R.\ Balasubramanian, B.\ S.\ Sathyaprakash, and S.\ V.\ Dhurandhar,
Phys.\ Rev.\ D {\bf 53}, 3033 (1996). 

\bibitem{SD91}
B.\ S.\ Sathyaprakash and S.\ V.\ Dhurandhar,
Phys.\ Rev.\ D {\bf 44}, 3819 (1991).

\bibitem{Owen96}
B.\ J.\ Owen, Phys.\ Rev.\ D {\bf 53}, 6749 (1996); 
B.\ J.\ Owen and B.\ S.\ Sathyaprakash, Phys.\ Rev.\ D {\bf 60}, 022002 (1999).

\bibitem{DIS2}
T.\ Damour, B.\ R.\ Iyer, and B.\ S.\ Sathyaprakash,
Phys.\ Rev.\ D {\bf 62}, 084036 (2000).

\bibitem{PoissonAndWill} 
E.\ Poisson and C.\ M.\ Will, Phys.\ Rev.\ D {\bf 52,} 848 (1995).

\bibitem{BSS00}
B.\ S.\  Sathyaprakash, 
Class.\ Quant.\ Grav.\ {\bf 17}, L157 (2000).

\bibitem{BSS94}
B.\ S.\  Sathyaprakash, 
Phys.\ Rev.\ D {\bf 50}, R7111 (1994).
\end{thebibliography}
\end {document}